\documentclass[%
reprint,
 amsmath,amssymb,
 aps,
]{revtex4-2}
\usepackage{graphicx}
\usepackage{dcolumn}
\usepackage{bm}
\usepackage{color}

\begin{document}

\preprint{APS/123-QED}

\title{Spin alignment of vector mesons from quark dynamics in a rotating medium}
%

\author{Minghua Wei}
\email{weiminghua@fudan.edu.cn}
 \affiliation{Institute of Modern Physics, Fudan University, Shanghai 200433, China}
\author{Mei Huang}%
 \email{huangmei@ucas.ac.cn}
\affiliation{%
 School of Nuclear Science and Technology, University of Chinese Academy of Sciences, Beijing 100049, China\\
}%

\date{\today}

\begin{abstract}
Vorticities in heavy-ion collisions (HICs) are supposed to induce spin alignment and polarization phenomena of quarks and mesons. In this work,  we analyze the spin alignment of vector mesons $\phi$ and $\rho$ induced by rotation from quark dynamics in the framework of the Nambu-Jona-Lasinio (NJL) model. The rotating angular velocity induces mass splitting of spin components for vector $\phi,\rho$ mesons $M_{\phi,\rho}(\Omega)\simeq M_{\phi,\rho}(\Omega=0)-s_{z}\Omega$. This behavior contributes to the spin alignment of vector mesons $\phi,\rho$ in an equilibrium medium and naturally explains the negative deviation of $\rho_{00}-1/3$ for vector mesons. 
Incidentally, the positive deviation of $\rho_{00}-1/3$ under the magnetic field can also be easily understood from quark dynamics.
\end{abstract}

\maketitle


\section{Introduction}
\label{sec:Introductio}
Relativistic heavy-ion collision experiments provide a platform for studying quantum chromodynamics (QCD) matter under extreme conditions. It has been expected that quark-gluon plasma (QGP) can be created through heavy-ion collisions (HICs)~\cite{BRAHMS:2004adc}. In a specific case, a non-central collision, QGP is created with a large orbital angular momentum (OAM) at a range of $10^{4}-10^{5}\hbar$~\cite{Becattini:2007sr,Jiang:2016woz,Deng:2016gyh}. Meanwhile, a strong magnetic field can reach the magnitude of $10 m_{\pi}^{2}$ at the initial time of the HICs and evolves with time~\cite{Kharzeev:2007jp,Skokov:2009qp,Deng:2012pc}. The magnitude of the magnetic field decays severely while the averaged angular velocity still maintains its magnitude along the axis which is perpendicular to the reaction plane~\cite{Jiang:2016woz}. Therefore, the effects of a rotating medium will play significant roles in the QCD phase diagram, dilepton production rate and spin polarization. 

Spin alignment of vector meson $\phi$ and $K^{*0}$ has been one of the intriguing topics in HICs. Experimental evidence suggests that spin density matrix element $\rho_{00}$ has a remarkable deviation from $1/3$~\cite{ALICE:2019aid,STAR:2022fan}. ALICE collaboration has measured $\rho_{00}$ for $K^{*0}$ and $\phi$ meson at $\sqrt{s_{NN}}=2.76$ TeV in Pb-Pb collisions~\cite{ALICE:2019aid}, and a negative deviation from $1/3$ at lower transverse momentum has been reported. STAR collaboration had measured $\rho_{00}$ between $\sqrt{s_{NN}}=11.5$ and $200$ GeV in Au-Au collisions~\cite{STAR:2022fan}. In this collision energy range, a positive deviation from $1/3$ is reported for $\rho_{00}$ of the vector meson $\phi$, which can be explained by a $\phi$-meson field \cite{Sheng:2019kmk}. 

In fact, spin alignment is a multifactorial phenomenon in heavy ion collisions. Theoretical researchers have developed a quark coalescence model to explain the contribution from the vorticity and magnetic field ~\cite{Yang:2017sdk,Sheng:2020ghv}. Furthermore, other physical mechanisms also contribute to the spin alignment, such as turbulent color fields\cite{Muller:2021hpe}, local vorticity\cite{Xia:2020tyd}, glasma fields\cite{Kumar:2023ghs}, and gradient corrections in the medium for vector mesons\cite {Li:2022vmb,Wagner:2022gza}.

Both the magnetic field and vorticities are expected to contribute to the spin alignment. In Ref.\cite{Sheng:2022ssp}, the effect of the magnetic field on spin alignment has been studied. In this work, we introduce the effects of global rotation on quark matter. Generally, rotation has an inhibiting effect on quark and anti-quark pairing~\cite{Jiang:2016wvv}, which is different from Magnetic Catalysis (MC)~\cite{Klevansky:1989vi,Klimenko:1991he,Gusynin:1995nb,Miransky:2015ava}. As a consequence, chiral condensate will disappear as the angular velocity grows. Since the chiral condensate is an order parameter for the chiral phase transition, its behavior demonstrates that a first-order phase transition occurs in low temperature regions and a crossover occurs in high temperature regions. The characteristics imply an analogy between the rotation and chemical potential~\cite{Chen:2015hfc,Chen:2021aiq}. However, it is found that the rotation behaves beyond an effective chemical potential in dilepton production~\cite{Wei:2021dib}, and the rotation enhances the dilepton production rate and induces the ellipticity of lepton pairs.

It is also worth emphasizing that previous studies have assumed an infinite size without boundary conditions~\cite{Jiang:2016woz,Wang:2018sur,Chen:2015hfc}. The boundary conditions on a finite-size system should be taken into account for a rotating system\cite{Ebihara:2016fwa,Chernodub:2017ref,Chernodub:2020qah,Fujimoto:2021xix}, and inhomogeneous chiral condensation will be developed \cite{Wang:2018zrn,Chen:2022mhf}.
For example, no-flux or MIT bag boundary conditions are applied in a spherical or cylindrical system~\cite{Chernodub:2016kxh,Zhang:2020hha,Zhang:2020jux}. The choice of boundary conditions will influence the behavior of chiral condensate near the surface. Particularly, when a no-flux boundary condition is applied in a cylindrical system, chiral condensate will almost remain constant in the inner part of the cylinder but get enhanced like a Gibbs phenomenon near the surface~\cite{Ebihara:2016fwa}. 

 In QCD phase diagram studies, it is also possible to consider inhomogeneous rotation with boundary conditions. Chiral condensate is enhanced near the area where the angular velocity changes severely, and this phenomenon is called the centrifugal effect~\cite{Wang:2018zrn}. Furthermore, chiral condensate reveals the dynamical mass of quarks. Besides the light quarks $u$ and $d$, heavy flavor quarks, like $s$ and $c$, are studied in electromagnetic and rotational fields ~\cite{Chen:2020xsr}. Corresponding $\phi, D$ mesons are expected to be influenced by rotation as well. However, it is complicated to describe the freeze-out of heavy particles from the QGP~\cite{He:2011qa}. Therefore, we only consider light-flavor vector meson modes excited and annihilated in a rotating medium. 


External fields, such as magnetic field and vortical field,  will induce the mass splitting of spin-1 vector mesons. In recent years, much attention has been paid to the magnetic field effect on vector mesons, such as $\rho,\phi$. For example, in case of charged $\rho$ mesons, mass splitting behaves like:
\begin{equation}
    M_{\rho^{\pm}}(eB) = M_{\rho^{\pm}}(0)\mp \kappa eB
\end{equation}
where $\kappa$ is a coefficient. For a point-like particle, $\kappa=1$; when the quark-antiquark loop effect is considered, $\kappa\neq 1$. Similarly, $\rho$ meson under rotation also exhibits a mass splitting phenomenon \cite{WeiMingHua:2020eee}:
\begin{equation}
    M_{\rho}(\Omega)=M_{\rho}(0)-s_{z}\Omega
\end{equation}
This relation indicates that vector mesons tend to occupy the $s_{z}=+1$ state. In this paper, a similar result can be obtained for the vector meson $\phi$. 

Recently, the mass splitting and spin alignment of vector meson $\phi$ have been investigated under a magnetic field~\cite{Sheng:2022ssp}. Another important contribution from the rotation on the spin alignment should also be taken into account. Therefore, in this work, we 
calculate the rotational contribution on matrix element $\rho_{00}$.

This paper is organized as follows. In section~\ref{sec:formalism}, quantum field theory in a rotating frame is introduced. Based on it, the quark propagator and the self-energy are modified by uniform rotation. Consequently, masses of vector mesons are obtained by random phase approximation. In section~\ref{sec:numericalresult}, the numerical results of quark mass, meson mass and spin alignment will be presented. Finally, the conclusion and summary will be given in section~\ref{sec:conclusion}. 

\section{Formalism}
\label{sec:formalism}
\subsection{Quantum field theory in a uniformly rotating frame}
\label{subsec:rotframe}
In quantum field theory, a Lorentz vector field $V^{a}(x)$ can be described by tetrad $e_{a}(x)=e_{a}^{\mu}(x)\frac{\partial}{\partial x^{\mu}}$ in curved space-time. In this section, Latin letters, $a,b=0,1,2,3$, stand for Lorentz indices. Greek letters, $\mu,\nu=0,1,2,3$, stand for coordinate indices. Parallel transport of $V^{a}(x)$ acts as follows:
\begin{equation}
    V^{a}(x+dx)=V^{a}(x)+\Gamma_{ab\mu}V^{b}dx^{\mu},
\end{equation}
where $\Gamma_{ab\mu}$ is called spin connection. Metric compatibility ensures $\Gamma_{ab\mu}=-\Gamma_{ba\mu}$, and the non-torsion condition provides a relation between the tetrad and the spin connection:
\begin{equation}\label{eq:spinconnection}
\begin{gathered}
\Gamma_\mu^{a b}=\frac{1}{2}\left[e^{a \nu}\left(\partial_\mu e_\nu^b-\partial_\nu e_\mu^b\right)-e^{b \mu}\left(\partial_\mu e_\nu^a-\partial_\nu e_\mu^a\right)\right. \\
\left.-e^{a \rho} e^{b l}\left(\partial_\rho e_{c \sigma}-\partial_\sigma e_{c \rho}\right) e_\mu^c\right].
\end{gathered}
\end{equation}
For spinor field, corresponding spinor connection is given by $\Gamma_{\mu}=\sigma^{ab}\Gamma_{ab\mu}$, where $\sigma^{ab}$ is spinor representation of Lorentz group. In the co-moving frame of the QGP, free fermions are described by the modified Dirac equation:
\begin{equation}
    [i\bar{\gamma}^{\mu}(\partial_{\mu}+\Gamma_{\mu})-M_{f}]\psi=0,
\end{equation}
where $M_{f}$ is the mass of a fermion and $\bar{\gamma}^\mu=e_a^\mu \gamma^a$ satisfies $\{\bar{\gamma}_{\mu},\bar{\gamma}_{\nu}\}=g^{\mu\nu}$. 

Particularly, in a uniformly rotating frame, the tetrad can be described by 
\begin{equation}
\label{eq:tetard}
e_\mu^a=\delta_\mu^a+\delta_i^a \delta_\mu^0 v_i, \quad e_a^\mu=\delta_a^\mu-\delta_a^0 \delta_i^\mu v_i,
\end{equation}
where $\vec{v}=\vec{\Omega} \times \vec{x}$ gives the velocity in the lab frame. We use the capital Greek letter $\Omega$ to represent the magnitude of angular velocity. If we substitute Eq.\eqref{eq:tetard} into Eq.\eqref{eq:spinconnection}, non-zero terms of the spin connection will be expressed as follows:
\begin{equation}
\begin{aligned}
\Gamma_{i j 0}=\frac{1}{2}\left(\partial_i v_j-\partial_j v_i\right), & \Gamma_{i 0 j}=\frac{1}{2}\left(\partial_i v_j+\partial_j v_i\right), \\
\Gamma_{0 i j}=-\frac{1}{2}\left(\partial_i v_j+\partial_j v_i\right), & \Gamma_{0 i 0}=-\frac{1}{2}\left(v_j \partial_i v_j+v_j \partial_j v_i\right) .
\end{aligned}
\end{equation}
Finally, in a uniformly rotating frame, the Dirac equation can be rewritten by\cite{Vilenkin:1980zv}
\begin{equation}
\label{eq:diracequation}
    [i\gamma^{a}\partial_{a}+\gamma^{0}\Omega \hat{J_{z}}-M_{f}]\psi=0.
\end{equation}
Here, the $z$-axis is chosen as the direction of the rotation, and it is perpendicular to the reaction plane. The total angular momentum $\hat{J_{z}}$ can be expressed as $\hat{J}_{z}=\hat{L}_{z}+\hat{S}_{z}$ where $\hat{L}_{z}$ is the orbital angular momentum and  $\hat{S}_{z}=\frac{1}{2}\left(
                      \begin{array}{cc}
                        \hat{\sigma}_{z} & 0 \\
                        0 & \hat{\sigma}_{z} \\
                      \end{array}
                    \right)$ contributes to the spin part.

For solving modified Dirac equation Eq.\eqref{eq:diracequation}, Ref.\cite{Jiang:2016wvv} has given the solutions in cylindrical coordinates where a position in space-time is labeled by $\tilde{r}=(t, r, \theta, z)$. We can write down a complete set of commuting operators: the Hamiltonian $\hat{H}$, the longitudinal momentum $\hat{k}_z$, the square of transverse momentum $\hat{\vec{k}}_t^2$, the total angular momentum $\hat{J}_z$, and the helicity operator on transverse plan $\hat{h}_t$. The eigenstates for fermion and anti-fermion  \cite{Jiang:2016wvv} are given by:
\begin{equation}
\label{eq:statesfermion}
U=\sqrt{\frac{E_k+M_{f}}{4 E_k}} e^{i k_z z} e^{i n \theta}\left(\begin{array}{c}
J_n\left(k_t r\right) \\
s e^{i \theta} J_{n+1}\left(k_t r\right) \\
\frac{k_z-i s k_t}{E_k+M_{f}} J_n\left(k_t r\right) \\
\frac{-s k_z+i k_t}{E_k+M_{f}} e^{i \theta} J_{n+1}\left(k_t r\right)
\end{array}\right), \\
\end{equation}
\begin{equation}
\label{eq:statesantifermion}
V=\sqrt{\frac{E_k+M_{f}}{4 E_k}} e^{-i k_z z} e^{i n \theta}\left(\begin{array}{c}
\frac{k_z-i s k_t}{E_k+M_{f}} J_n\left(k_t r\right) \\
\frac{s k_z-i k_t}{E_k+M_{f}} e^{i \theta} J_{n+1}\left(k_t r\right) \\
J_n\left(k_t r\right) \\
-s e^{i \theta} J_{n+1}\left(k_t r\right)
\end{array}\right),
\end{equation}
where $J_n\left(k_t r\right)$ is the $n$-th order Bessel function and $n\in \mathbb{Z}$ stands for the quantum number of angular momentum. In Eq.\eqref{eq:statesfermion} and Eq.\eqref{eq:statesantifermion}, $k_{t}$ and $k_{z}$ are the eigenvalues of transverse and longitudinal momentum, respectively, and $s=\pm 1$ is the eigenvalue of the helicity operator $\hat{h}_t$. Besides, $E_{k}$ is defined by $E_k \equiv \sqrt{k_z^2+k_t^2+M_{f}^2}$ and the energy is given by $E=\pm E_k-(n+1/2) \Omega$.

Based on the solutions of Dirac equation, one can write down the quark propagator by definition: $S\left(\tilde{r};\tilde{r}^{\prime}\right)=\left\langle 0\left|T\psi(\tilde{r})\bar{\psi}\left(\tilde{r}^{\prime}\right)\right| 0\right\rangle$. As a standard procedure, $\psi(\tilde{r})$ and $\bar{\psi}\left(\tilde{r}^{\prime}\right)$ are expanded by Eq.\eqref{eq:statesfermion} and Eq.\eqref{eq:statesantifermion}, and a explicit form can be obtained as following:
\begin{widetext}
\begin{equation}
\label{eq:propagator}
\begin{aligned}
S\left(\tilde{r} ; \tilde{r}^{\prime}\right)= & \frac{1}{(2 \pi)^2} \sum_n \int \frac{d k_0}{2 \pi} \int k_t d k_t \int d k_z \frac{e^{i n\left(\theta-\theta^{\prime}\right)} e^{-i k_0\left(t-t^{\prime}\right)+i k_z\left(z-z^{\prime}\right)}}{\left[k_0+\left(n+\frac{1}{2}\right) \Omega\right]^2-k_t^2-k_z^2-M_f^2+i \epsilon} \\
& \times\left\{\left[\left[k_0+\left(n+\frac{1}{2}\right) \Omega\right] \gamma^0-k_z \gamma^3+M_f\right]\left[J_n\left(k_t r\right) J_n\left(k_t r^{\prime}\right) \mathcal{P}_{+}+e^{i\left(\theta-\theta^{\prime}\right)} J_{n+1}\left(k_t r\right) J_{n+1}\left(k_t r^{\prime}\right) \mathcal{P}_{-}\right]\right. \\
& \left.-i \gamma^1 k_t e^{i \theta} J_{n+1}\left(k_t r\right) J_n\left(k_t r^{\prime}\right) \mathcal{P}_{+}-\gamma^2 k_t e^{-i \theta^{\prime}}J_n\left(k_t r\right) J_{n+1}\left(k_t r^{\prime}\right) \mathcal{P}_{-}\right\}.
\end{aligned}
\end{equation}
\end{widetext}
Here, we have simplified the expression by projection operators $\mathcal{P}_{\pm}=\frac{1}{2}\left(1 \pm i \gamma^1 \gamma^2\right)$. In Appendix \ref{sec:appendixa}, we will present another procedure for obtaining the quark propagator in a rotating medium. 

Nevertheless, Eq.\eqref{eq:propagator} is the result under the infinite-size approximation. One can find a finite-size version with a boundary condition (see Ref.\cite{Ebihara:2016fwa}). The transverse momentum $k_{t}$ is discrete, and its integral is replaced by the summation of a series. In this manuscript, firstly, we have to calculate the spectral functions of vector mesons in a rotating medium. So, we apply the infinite-size approximation to avoid the tedious summation of the series. This compromise will cause the violation of causality in the case of large angular velocity. In this manuscript, we study the phenomena and quantities at a fixed radius $r=0.1 \text{GeV}^{-1}$ and in a range of angular velocity from $\Omega=0$ GeV to $\Omega= 1.2 $ GeV so that the velocity of a fixed point is smaller than the speed of light, i.e. $\Omega r < 1$.

\subsection{The 3-flavor NJL model}
In order to study the microscopic properties of vector mesons, we use the Nambu-Jona-Lasinio(NJL) model to estimate the strong interaction\cite{Klevansky:1992qe}. The 3-flavor NJL model is required to investigate the $\phi$ meson, which contains an $s$ quark \cite{Fukushima:2008wg}. The Lagrangian of the 3-flavor NJL model is given as follows:
\begin{equation}
\begin{aligned}
\mathcal{L}_{3 \mathrm{NJL}}=& \bar{\psi}[i\bar{\gamma}^{\mu}(\partial_{\mu}+\Gamma_{\mu})-m_{f}]\psi\\
&+G_{S} \sum_{a=0}^8\left[\left(\bar{\psi} \lambda^a \psi\right)^2+\left(\bar{\psi} i \gamma_5 \lambda^a \psi\right)^2\right] \\
&-G_{V} \sum_{a=0}^8\left[\left(\bar{\psi} \gamma_{\mu}\lambda^a \psi\right)^2+\left(\bar{\psi} i \gamma_{\mu}\gamma_5 \lambda^a \psi\right)^2\right] \\
&-K\left[\operatorname{det} \bar{\psi}\left(1+\gamma_5\right) \psi+\operatorname{det} \bar{\psi}\left(1-\gamma_5\right) \psi\right],
\end{aligned}
\end{equation}
where $K$ is the coupling constant of the six-fermion interaction, $G_{s}$ and $G_{V}$ are coupling constants of the four-fermion interaction for scalar and vector channels, respectively. In the 3-flavor NJL model, $\psi=(\psi_{u},\psi_{d},\psi_{s})$ is a Dirac spinor which contains $u$, $d$ and $s$ quarks. Correspondingly, $\lambda_{a}$ are the Gell-Mann matrices and $m_{f}$ is the current quark mass with different flavors.

The NJL model is an effective model that only contains quarks. The rotation affects the quark dynamics through the non-zero term of the spinor connection $\Gamma^{\mu}$. Furthermore, by applying the mean field approximation, the Lagrangian will be rewritten as:
\begin{equation}
\begin{aligned}
\mathcal{L}_{\mathrm{MF}}=& \sum_{f=u, d, s} \bar{\psi}_f\left[i \bar{\gamma}_\mu (\partial_{\mu}+\Gamma_{\mu})-M_f\right] \psi_f \\
&-2 G_S \sum_{f=u, d, s} \sigma_f^2+4 K \sigma_u \sigma_d \sigma_s
\end{aligned}
\end{equation}
where $\sigma_{f}$ is the chiral condensate $\sigma_f \equiv\left\langle\bar{\psi}_f \psi_f\right\rangle$ for a specific flavor. And $M_{f}$ is the dynamical quark mass, which is given by
\begin{equation}
\label{eq:dynamicalquarkmass}
M_f \equiv m_f-4 G_S \sigma_f+2 K \prod_{f^{\prime} \neq f} \sigma_{f^{\prime}}.
\end{equation}
In a uniformly rotating medium, the spinor connection $\Gamma^{\mu}$ has been estimated in section \ref{subsec:rotframe}. We assume the direction of rotation is parallel to the z-axis. By applying the standard procedure in finite-temperature field theory \cite{kapusta_gale_2006}, the grand potential for quarks with specific flavor is shown as follows:
\begin{equation}
\begin{aligned}
\Omega_{f}(r)=&\frac{N_c}{8 \pi^2} T \sum_n \int d k_t^2 \int d k_z\left[J_n\left(k_t r\right)^2+J_{n+1}\left(k_t r\right)^2\right] \\
&\times\left[E_{k}/T+\ln \left(1+e^{-\left(E_k-\left(n+\frac{1}{2}\right) \Omega\right) / T}\right)\right. \\
&\left.+\ln \left(1+e^{-\left(E_k+\left(n+\frac{1}{2}\right) \Omega\right) / T}\right)\right].
\end{aligned}
\end{equation}
Consequently, the total grand potential is
\begin{equation}
\Omega_{\text{tot}}(r)=\sum_{f=u, d, s}\left(2 G_S \sigma_f^2-\Omega_f\right)+4 K \sigma_u \sigma_d \sigma_s.
\end{equation}
Here, $N_{c}=3$ is the degeneracy of color and $T$ is the temperature of the medium. We can obtain the dynamical quark mass $M_{f}$ and chiral condensate $\sigma_{f}$ by solving the gap equations:
\begin{equation}
\frac{\partial \Omega_{\text{tot}} }{ \partial \sigma_f}=0,\hspace{5pt}
\frac{\partial^{2} \Omega_{\text{tot}} }{ \partial \sigma_f^{2}}>0.
\end{equation}
In principle, if other parameters are fixed, the dynamical quark mass $M_{f}(r)$ will be a function of radius $r$. Currently, most of studies have assumed that $M_{f}(r)$ is changed smoothly, i.e. $\partial M_{f}(r)/\partial r\simeq 0$. Such assumption is called local density approximation(LDA)\cite{Chen:2015hfc,Jiang:2016wvv,Wang:2018zrn}. In this manuscript, we use the LDA and choose a fixed radius $r=0.1\text{GeV}^{-1}$. Then, $M_{f}(\Omega)$ will be evaluated numerically in section \ref{sec:quarkmass}.

\subsection{Vector meson mass splitting under the rotation}

In the NJL model, the vector meson $\phi$ is treated as an $s\bar{s}$ bound state or a resonance state, and it can be constructed by quark-antiquark scattering\cite{Klevansky:1992qe}. In the random phase approximation (RPA), the vector meson propagator can be obtained by summation of quark loops and the one-loop polarization function is given by
\begin{equation}
\label{eq:polfun}
\Pi^{\mu\nu}(q)=-i \int d^{4}\tilde{r}\text{Tr}_{sfc}[i \gamma^{\mu}S(0;\tilde{r})i \gamma^{\nu}S(\tilde{r};0)]e^{i q\cdot \tilde{r}}.
\end{equation}
Here, $\text{Tr}_{sfc}$ stands for the trace in the spinor, flavor and color spaces. Since the $\phi$ meson is purely constituted by $s$ quarks, $S(0;\tilde{r})$ stands for the $s$ quark propagator with the dynamical mass $M_{s}$ given by the mean field approximation. The polarization function of the $\phi$ meson is supposed to be modified in a rotating medium. 

Currently, we focus on the mesons which remain at rest in the rotating frame, i.e., $\vec{q}=0$. In this case, polarization vectors are given by
\begin{equation}
\begin{aligned}
\epsilon^{\mu}_{1}&=\frac{1}{\sqrt{2}}(0,1,i,0),\\ \epsilon^{\mu}_{2}&=\frac{1}{\sqrt{2}}(0,1,-i,0),\\
b^{\mu}&=(0,0,0,1)
\end{aligned}
\end{equation}
where $\epsilon^{\mu}_{1}$ and $\epsilon^{\mu}_{2}$ are the right and left-hand polarization vectors respectively. The longitudinal polarization vector is parallel to the direction of rotation. Correspondingly, the projection operators are given by
\begin{equation}
\label{eq:projectionoperation}
\begin{aligned}
P^{\mu\nu}_{1}&=-\epsilon^{\mu}_{1}\epsilon^{\nu}_{1},(s_{z}=-1 \text{ for } \phi \text{ meson }),\\
P^{\mu\nu}_{2}&=-\epsilon^{\mu}_{2}\epsilon^{\nu}_{2},(s_{z}=+1\text{ for } \phi \text{ meson }),\\
L^{\mu\nu}&=-b^{\mu}b^{\nu},(s_{z}=0 \text{ for } \phi \text{ meson }).
\end{aligned}
\end{equation}

As we have assumed $\vec{q}=0$, nonzero elements of the polarization functions can be read in the following matrix:
\begin{equation}
\Pi^{\mu\nu}_{\phi}=\left(
                      \begin{array}{cccc}
                        0 & 0 & 0 & 0 \\
                        0 & \Pi^{11} & \Pi^{12} & 0 \\
                        0 & \Pi^{21} & \Pi^{22} & 0 \\
                        0 & 0 & 0 & \Pi^{33} \\
                      \end{array}
                    \right).
\end{equation}
The explicit expressions of matrix elements are shown in Ref.\cite{Wei:2021dib}. This tensor can be decomposed by projection operators in Eq.\eqref{eq:projectionoperation} as follows:
\begin{equation}
\Pi^{\mu\nu}_{\phi}=A_{-1,-1}P^{\mu\nu}_{1}+A_{11}P^{\mu\nu}_{2}+A_{00}L^{\mu\nu},
\end{equation}
Similarly, the vector meson propagator can be decomposed as follows:
\begin{equation}
D^{\mu\nu}_{\phi}(q)=D_{-1,-1}(q)P^{\mu\nu}_{1}+D_{11}(q)P^{\mu\nu}_{2}+D_{00}(q)L^{\mu\nu}.
\end{equation}
Here, coefficients $D_{\lambda}$ are obtained by RPA summation and expressed by:
\begin{equation}
D_{\lambda\lambda}(q)=\frac{4G_{V}}{1+4G_{V}A_{\lambda\lambda}},
\end{equation}
where
\begin{equation}
\label{coeff}
\begin{aligned}
A_{-1,-1}&=-(\Pi_{11} - i \Pi_{12}), &(s_{z}=-1 \text{ for } \phi \text{ meson }),\\
A_{11}&=-\Pi_{11} - i \Pi_{12}, &(s_{z}=+1\text{ for } \phi \text{ meson }),\\
A_{00}&=-\Pi_{33}, &(s_{z}=0 \text{ for } \phi \text{ meson }).
\end{aligned}
\end{equation}
Then we can obtain the corresponding spectrum functions for different spin components, which take the following form:
\begin{equation}
\begin{aligned}
\xi_{\lambda\lambda}(\omega) &\equiv \frac{1}{\pi} \operatorname{Im} D_{\lambda\lambda}(\omega)\\
&=\frac{\left(4 G_V\right)^2 \operatorname{Im} A_{\lambda\lambda}(\omega)}{\pi\left\{\left[1+4 G_V \operatorname{Re} A_{\lambda\lambda}(\omega)\right]^2+\left[4 G_V \operatorname{Im} A_{\lambda\lambda}(\omega)\right]^2\right\}},
\label{Eq:spectrafunction}
\end{aligned}
\end{equation}
where $\omega$ is the energy of the meson and we set the momentum to $\vec{q}=0$.

\subsection{spin alignment of vector meson $\phi$}
Heavy-ion collisions will create an ensemble of particles under extreme conditions. Particularly, we consider vector meson $\phi$ with spin-1. For a fixed direction, a normalized spin state of a $\phi$ meson is labeled by $\left|\lambda\right\rangle$ with $\lambda=1,0,-1$. Spin density operator $\rho$ is defined by
\begin{equation}
\rho=\sum_{\lambda\lambda^{\prime}} \rho_{\lambda\lambda^{\prime}}\left|\lambda\right\rangle\left\langle\lambda^{\prime}\right|,
\end{equation}
Here, $\rho_{\lambda\lambda^{\prime}}$ comprises a $3\times 3$ spin density matrix as following:
\begin{equation}
\rho_{\lambda\lambda^{\prime}}=\left(\begin{array}{ccc}
\rho_{11} & \rho_{10} & \rho_{1,-1}\\
\rho_{01} & \rho_{00} & \rho_{1,-1}\\
\rho_{-1,1}& \rho_{-1,0} & \rho_{-1,-1}
\end{array}\right)\,,
\end{equation}
An ensemble of $\phi$ mesons, for which spin information is described by $\rho_{\lambda\lambda^{\prime}}$, will decay to 
\begin{equation}
\phi \rightarrow K^{+}+K^{-}.
\end{equation} 
In this process, daughter particles will have an angular distribution\cite{Schilling:1969um,Liang:2004xn}:
\begin{equation}
\begin{aligned}
\frac{d N}{d \Omega^{*}}= & \frac{3}{4 \pi}\left\{\cos ^2 \theta \rho_{00}+\sin ^2 \theta\left(\rho_{11}+\rho_{-1-1}\right) / 2\right. \\
& -\sin 2 \theta\left(\cos \phi \operatorname{Re} \rho_{10}-\sin \phi \operatorname{Im} \rho_{10}\right) / \sqrt{2} \\
& +\sin 2 \theta\left(\cos \phi \operatorname{Re} \rho_{-10}+\sin \phi \operatorname{Im} \rho_{-10}\right) / \sqrt{2} \\
& \left.-\sin ^2 \theta\left[\cos (2 \phi) \operatorname{Re} \rho_{1-1}-\sin (2 \phi) \operatorname{Im} \rho_{1-1}\right]\right\} .
\end{aligned}
\end{equation}
From experimental data of the $\theta$-distribution, $\rho_{00}$ can be obtained as a coefficient of angular distribution. Since experiments can measure the value of $\rho_{\lambda\lambda^{\prime}}$,
theoretical studies should explain the results of spin alignment and evaluate $\rho_{\lambda\lambda^{\prime}}$ qualitatively. Magnetic field and vorticity are two factors that are taken into account in many models. The approaches are generalized  as follows: 
\begin{equation}
\label{eq:aim}
\operatorname{B, \Omega}\left(\text{quarks}\right) \stackrel{\text { influence }}{\longrightarrow} \rho_{\lambda\lambda^{\prime}}\left( \phi\right) \stackrel{\text { determine }}{\longrightarrow} \frac{d N}{d \Omega^{*}}\left( K^{+},K^{-}\right)
\end{equation}
Theoretical studies are interested in the first step of Eq.\eqref{eq:aim}. In a real heavy-ion collision process, the evolution is complicated, and the history will influence $\rho_{\lambda\lambda^{\prime}}$ after the freeze-out. One possible method is the quark coalescence model. However, it is unable to take the medium effects into account in the quark coalescence model. As we mentioned in previous sections, quark mass and meson mass will be modified in a rotating medium.

In this investigation, we aim at a uniformly rotating medium, and the created $\phi$ mesons are in global equilibrium. In this case,
the particle number density $\overline{\rho}_{\lambda\lambda^{\prime}}({\bf q})$ can be expressed by
\begin{equation}
\overline{\rho}_{\lambda\lambda^{\prime}}({\bf q})=\int d\omega\,\frac{2\omega}{e^{\omega/T}-1}\xi_{\lambda\lambda^{\prime}}(\omega, \bf q)\,,\label{eq:density matrix}
\end{equation}
where $\xi_{\lambda\lambda^{\prime}}(\omega, \bf k)$ is the spectral function given in Eq.(\ref{Eq:spectrafunction}). At the one-loop level, the spectral function in a rotating medium is calculated in Ref.\cite{Wei:2021dib}, and we present it in Appendix \ref{sec:appB} for convenience. Generally, the spectral function is the imaginary part of the full propagator:
\begin{equation}
\xi_{\lambda\lambda^{\prime}}(q)\equiv\frac{1}{\pi}\text{Im}D_{\lambda\lambda^{\prime}}(q)\,.\label{eq:spectra function}
\end{equation}
For a general nonzero momentum ${\bf q}$, the matrix $\xi_{\lambda\lambda^{\prime}}(\omega, \bf q)$ and $\overline{\rho}_{\lambda\lambda^{\prime}}({\bf q})$ may have nonzero off-diagonal elements.
Particularly, if the quantization direction coincides with the rotation axis and the momentum ${\bf q}$ is parallel/antiparallel to the rotation axis, or ${\bf q}=0$, $\xi_{\lambda\lambda^{\prime}}(\omega, \bf q)$ and $\overline{\rho}_{\lambda\lambda^{\prime}}({\bf q})$ will become diagonal:
\begin{equation}
\overline{\rho}_{\lambda\lambda^{\prime}}({\bf q})=\left(\begin{array}{ccc}
\overline{\rho}_{11} & 0 & 0\\
0 & \overline{\rho}_{00} & 0\\
0 & 0 & \overline{\rho}_{-1,-1}
\end{array}\right)\,.
\end{equation}
Therefore, we divide the spectral function into two parts: a delta function part and a continuum part, i.e. 
\begin{equation}
\xi_{\lambda\lambda}(\omega, \mathbf{q})=\delta\left(\omega^2-M_{\phi, \lambda}^2\right)+\xi_{\lambda\lambda}^*(\omega,\bf q)
\end{equation}
where $\xi_{\lambda\lambda}^*(\omega,\bf q)$ is the continuum part of spectral function and $M_{\phi,\lambda}$ is the vector meson mass for different spin components. Correspondingly the particle number density $\bar{\rho}_{\lambda\lambda}({\bf q})$ takes the form of:
\begin{equation}
\bar{\rho}_{\lambda\lambda}({\bf q})=\frac{1}{\exp \left(M_{\phi,\lambda} / T\right)-1}+\int d \omega \frac{2 \omega \xi_{\lambda\lambda}^*(\omega, \bf q)}{\exp (\omega / T)-1}
\end{equation}
In experiment measurement, spin alignment is evaluated by the matrix element $\rho_{00}$ which can be expressed by:
\begin{equation}
\rho_{00}({\bf q})\equiv\frac{\overline{\rho}_{00}({\bf q})}{\sum_{\lambda=0,\pm1}\overline{\rho}_{\lambda\lambda}({\bf q})}\,.\label{eq:spin alignment}
\end{equation}
In the non-rotating case, $\phi$ mesons with different spin components have the same mass $M_{\phi}(s_{z}=0,\pm 1)$, which leads to $\rho_{00}=1/3$. At a finite angular velocity, the mass of the $\phi$ meson with the $s_{z}=+1$ component decreases linearly. Thus, $\phi$ mesons tend to occupy the $s_{z}=+1$ state. As a consequence, spin polarization and spin alignment can be obtained in our formalism. 

In this section, Eqs.~(\ref{eq:density matrix}-\ref{eq:spin alignment}) are momentum-dependent. However, in this investigation, we evaluate the spectral functions and spin alignment with ${\bf q}={\bf 0}$, which means vector mesons are at rest in the rotating frame. In general, Ref.\cite{Sheng:2022ssp} has revealed the spin density matrix for arbitrary measuring direction, which is characterized by Euler angles
$(\alpha,\,\beta,\,\gamma)$. The explicit form is: 
\begin{align}
 & \overline{\rho}_{\lambda\lambda^{\prime}}({\bf 0};\alpha,\beta,\gamma)\nonumber \\
 & \ \ =\sum_{\lambda_{1},\lambda_{2}}R_{\lambda\lambda_{1}}(\alpha,\beta,\gamma)\overline{\rho}_{\lambda_{1}\lambda_{2}}({\bf 0})R_{\lambda_{2}\lambda^{\prime}}^{-1}(\alpha,\beta,\gamma)\,.
\end{align}
Here, $R_{\lambda\lambda^{\prime}}(\alpha,\beta,\gamma)$ is the spin-1
representation of the rotation. As a result, the spin alignment only depends on Euler
angles $\beta$, and the explicit form is:
\begin{equation}
\rho_{00}({\bf 0};\alpha,\beta,\gamma)=\frac{\overline{\rho}_{00}({\bf 0})\cos^{2}\beta+\overline{\rho}_{11}({\bf 0})\sin^{2}\beta}{\bar{\rho}_{00}({\bf 0})+\bar{\rho}_{11}({\bf 0})+\bar{\rho}_{-1,-1}({\bf 0})}\,.
\end{equation}
In Section \ref{sec:spinalignmentresult}, numerical results will be presented for $\beta=0$, and the following formula is used:
\begin{equation}
\rho_{00}^{\Omega}({\bf 0})\equiv\frac{\bar{\rho}_{00}({\bf 0})}{\bar{\rho}_{00}({\bf 0})+\bar{\rho}_{11}({\bf 0})+\bar{\rho}_{-1,-1}({\bf 0})}\,.
\end{equation}

\section{Numerical result}
\label{sec:numericalresult}
Since a rotating system has broken the Lorentz symmetry, it is not necessary to use Pauli-Villars regularization. In fact, the Bessel function will not reflect the oscillation behavior in the cut-off energy scale $\Lambda$. Therefore, we choose a soft cut-off function:
\begin{equation}
    f_{\Lambda}=\frac{\Lambda^{2*10}}{\Lambda^{2*10}+p^{2*10}}
\end{equation}
where $\Lambda=620.411 \text{MeV}$ is the energy scale for cut-off. And the corresponding coupling constants are: $\mathrm{G_{S}}=\frac{1.710}{\Lambda^2}$, $G_{V}=0.67 * \frac{1.710}{\Lambda^2}$ and $K=\frac{12035}{1000 \Lambda^5}$. In this case, $m_{\pi}=140$ MeV and $m_{\phi}=1020$ MeV in vacuum. Current quark masses are: $m_{u}=m_{d}=5.5$ MeV, $m_{s}=135.433$ MeV.

\subsection{Chiral condensates and dynamical quark masses}
\label{sec:quarkmass}

\begin{figure}[t]
    \centering
     \includegraphics[width=7cm]{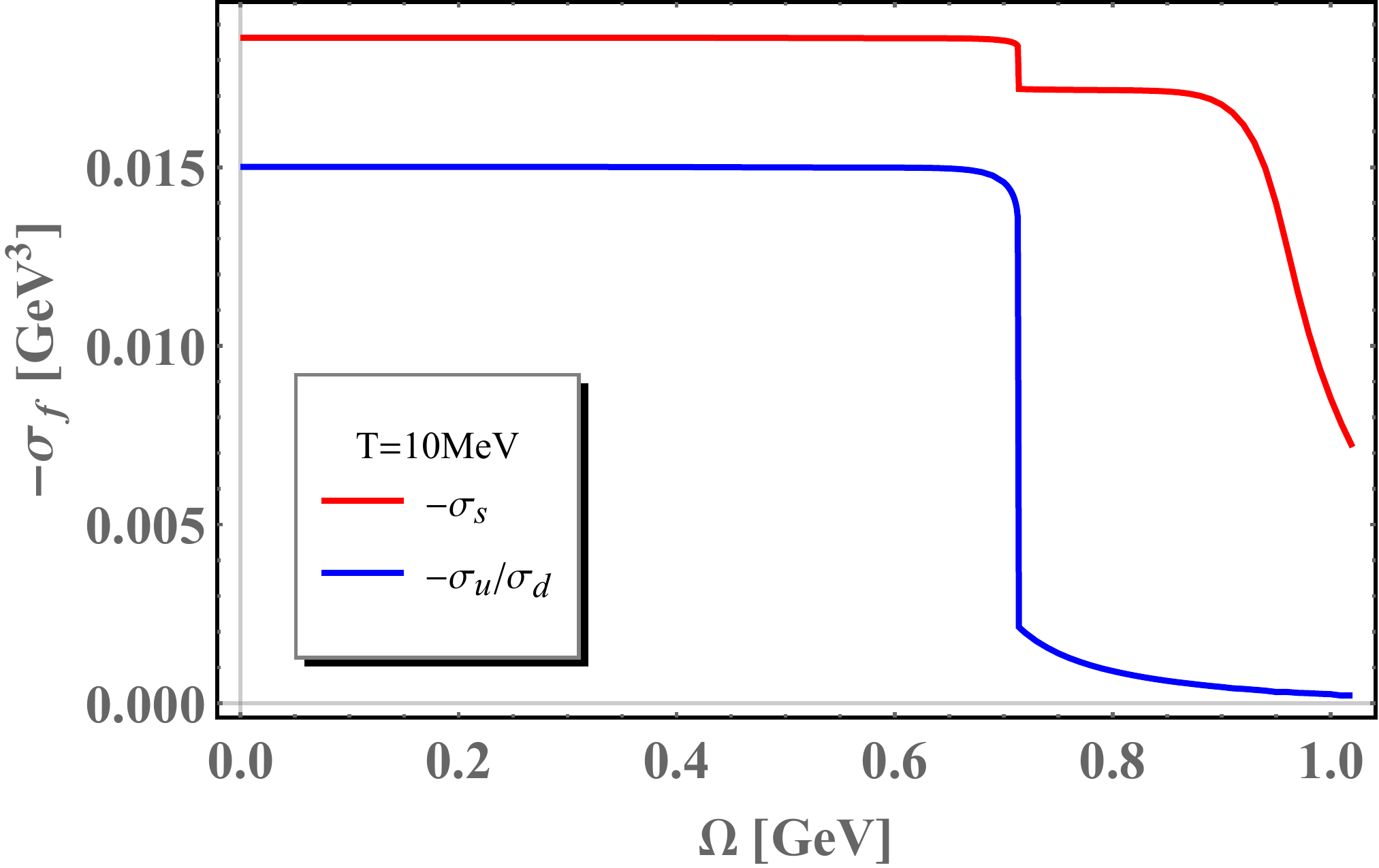}
    \caption{Chiral condensates as functions of angular velocity at temperature $T=10$ MeV. Red lines stand for $-\sigma_{s}$ and the blue lines stand for $-\sigma_{u}$ and $-\sigma_{d}$. }  
    \label{fig:condensate10}
\end{figure}

\begin{figure}[t]
    \centering
    \includegraphics[width=7cm]{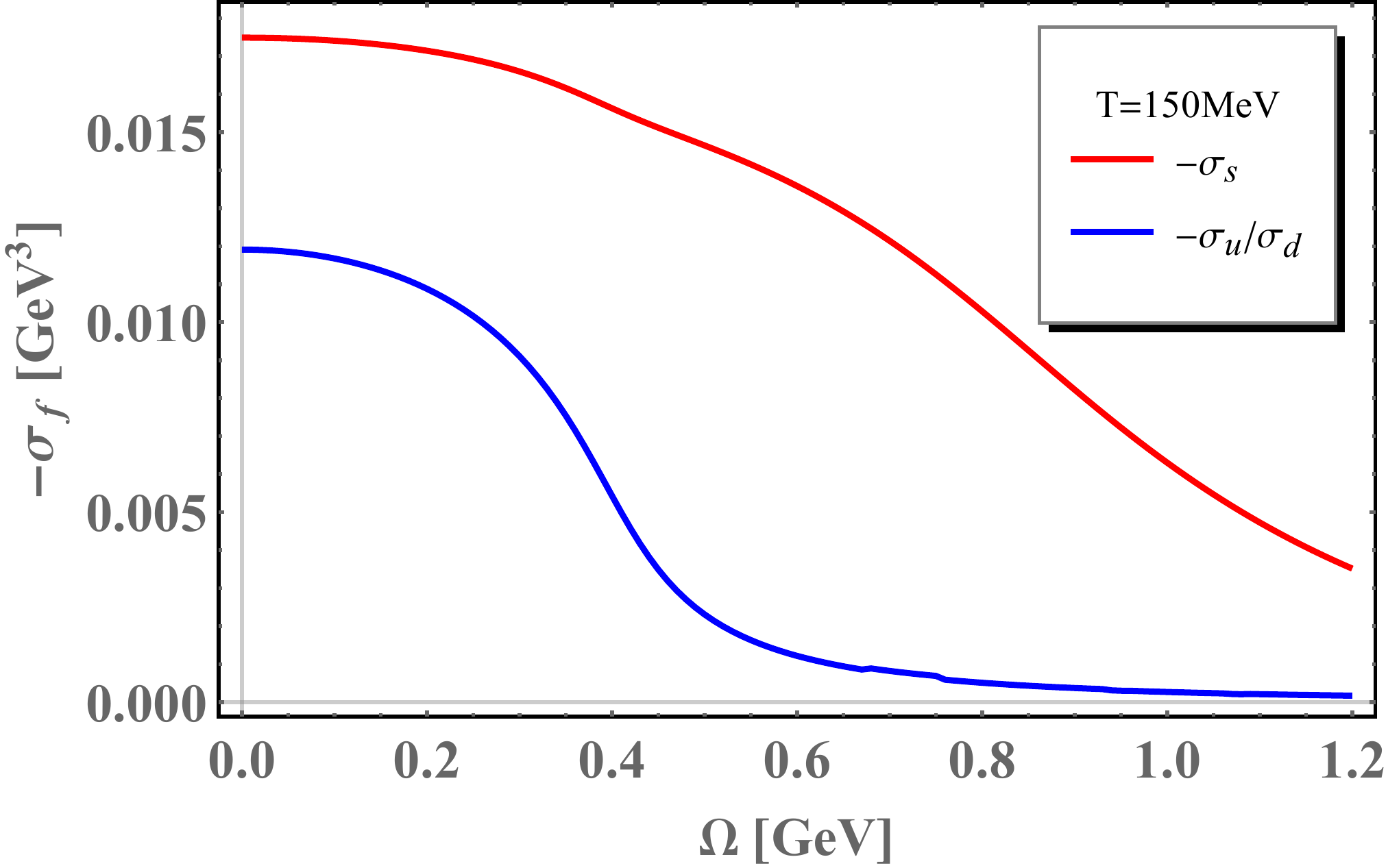}  \caption{Chiral condensates as functions of angular velocity at temperature $T=150$ MeV. Red lines stand for $-\sigma_{s}$ and the blue lines stand for $-\sigma_{u}$ and $-\sigma_{d}$. }  
    \label{fig:condensate150}
\end{figure}

\begin{figure}[t]
    \centering
     \includegraphics[width=7cm]{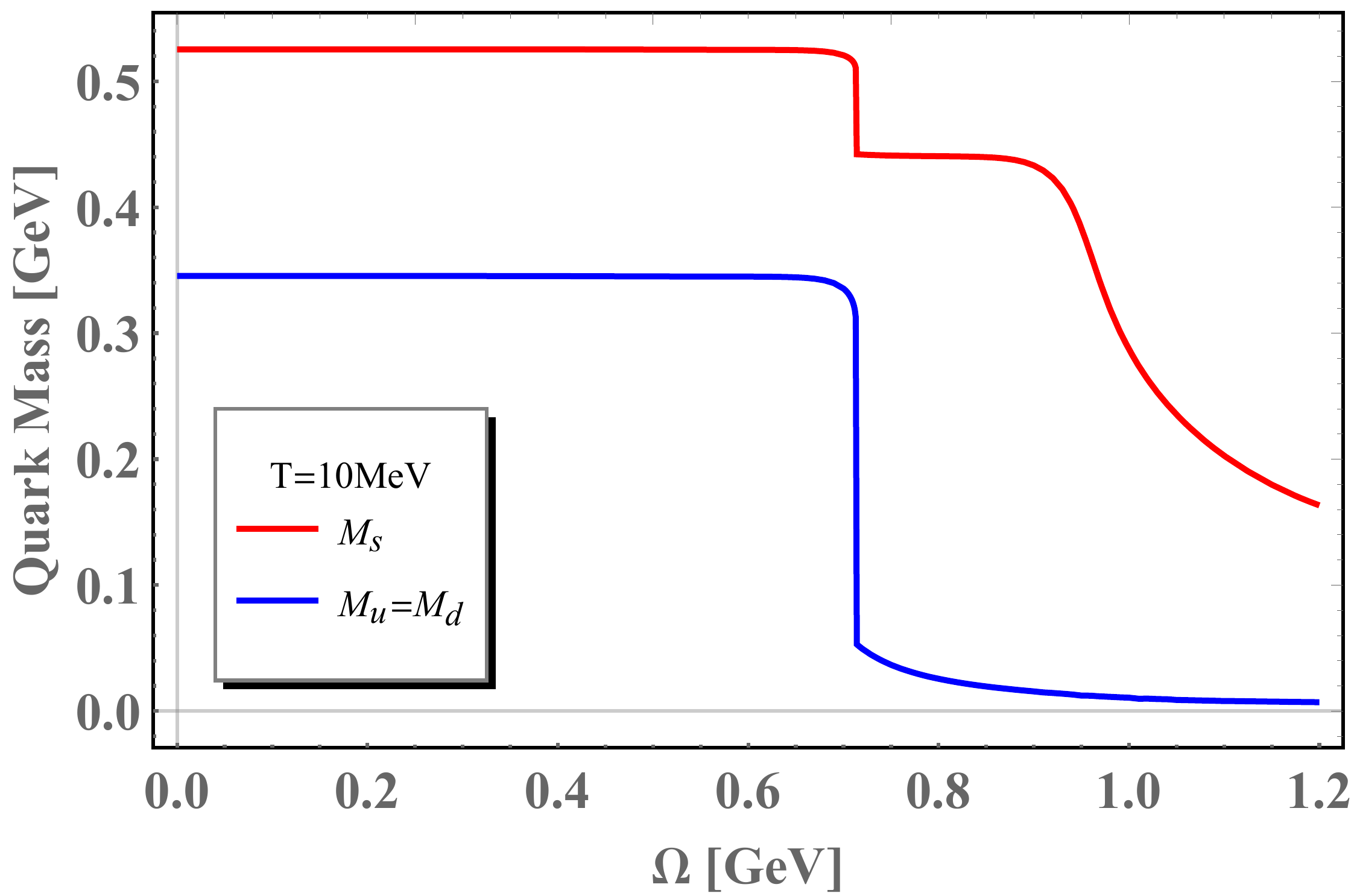}
    \caption{Dynamical quark masses as functions of angular velocity at temperature $T=10$ MeV. Red lines stand for the mass of the $s$ quark, and the blue lines stand for the light quarks $u$ and $d$. }  
    \label{fig:quarkmass10}
\end{figure}

\begin{figure}[t]
    \centering
    \includegraphics[width=7cm]{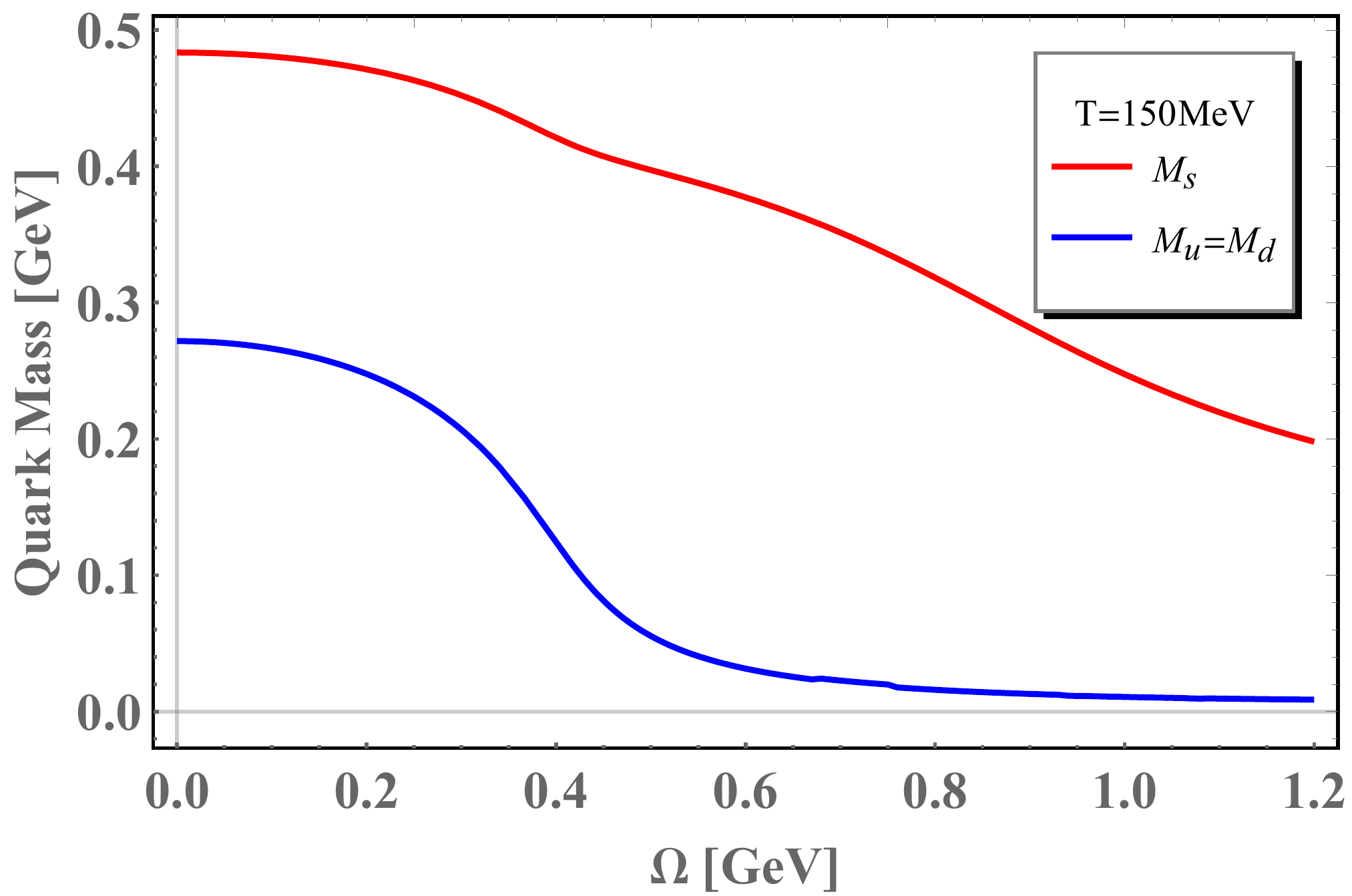}  \caption{Dynamical quark masses as functions of angular velocity at temperature $T=150$ MeV. Red lines stand for the mass of the $s$ quark and the blue lines stand for the light quarks $u$ and $d$. }  
    \label{fig:quarkmass150}
\end{figure}

As mentioned in previous sections, chiral condensate will be suppressed under rotation. In FIG. \ref{fig:condensate10} and FIG. \ref{fig:condensate150}, chiral condensates are demonstrated as a function of angular velocity $\Omega$ at temperature $T=10 \text{MeV}$ and $T=150 \text{MeV}$ respectively. Correspondingly, dynamical quark masses are also demonstrated as a function of angular velocity $\Omega$ in FIG. \ref{fig:quarkmass10} and FIG. \ref{fig:quarkmass150}. Chiral symmetry will be restored as angular velocity grows. In FIG. \ref{fig:condensate10} and FIG. \ref{fig:quarkmass10}, at an almost zero temperature, $T=10 \text{MeV}$, rotation induces a first order transition at angular velocity $\Omega_{c}=0.713$ GeV. Both $M_{u}$ and $M_{d}$ stay at a constant mass when the angular velocity is below $\Omega_{c}$. From Eq.\eqref{eq:dynamicalquarkmass}, we know that dynamical quark masses are determined by the chiral condensates $\sigma_{u}$, $\sigma_{d}$ and $\sigma_{s}$. In fact, $\sigma_{u}=\sigma_{d}$ and drops at $\Omega_{c}$ while $\sigma_{s}$ still varies smoothly. Consequently, $M_{s}$ has a small jump at $\Omega_{c}$ and then decreases smoothly. The behavior of $M_{u,d,s}$ as a function of angular velocity $\Omega$ is very similar to the case of finite quark chemical potential~\cite{Rehberg:1995kh,Buballa:2003qv}, but the first-order phase transition will occur at $\mu_c\simeq 0.33$ GeV at finite density. 

At a higher temperature, $T=150$ MeV, FIG.\ref{fig:quarkmass150} reveals that the chiral phase transition will be a crossover, which occurs around $\Omega_{c}\sim 0.4$ GeV. The phase transition takes place in a smaller angular velocity. Furthermore, it is noticed that the mass decreases slowly before the phase transition. Above all, the rotational effect on the dynamical quark mass is similar to that of the chemical potential $\mu_{q}$.

\subsection{Mass spectra of $\phi$ and $\rho$ meson under rotation}

After we obtain the dynamical quark masses at different temperatures and angular velocities, we can apply this result in Eq.(\ref{eq:polfun}) and obtain corresponding meson masses. 

At zero temperature and low temperatures, similar to $\rho$ meson, $\phi$ meson mass with different spin components $s_z=0,\pm1$ also show mass splitting effect with $M_{\phi}(\Omega)= M_{\phi}(0)-s_{z}\Omega$. 
\begin{figure}[t]
    \centering
    \includegraphics[width=7cm]{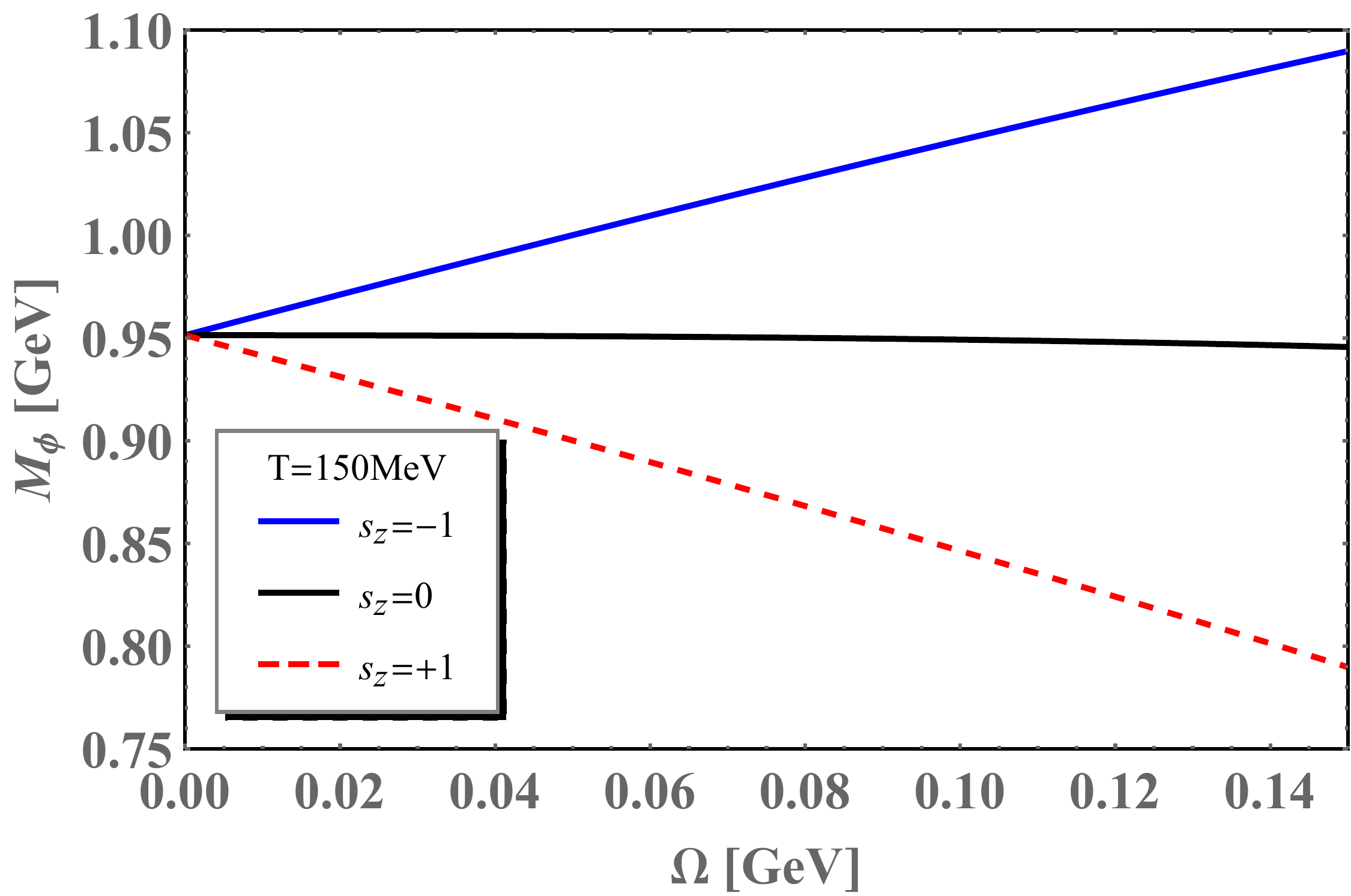}
    \caption{$\phi$ meson mass as a function of angular velocity at temperature $T=150$ MeV and $\mu=0$ MeV.}  
    \label{fig:masssplit}
\end{figure}
\begin{figure}[t]
    \centering
     \includegraphics[width=7cm]{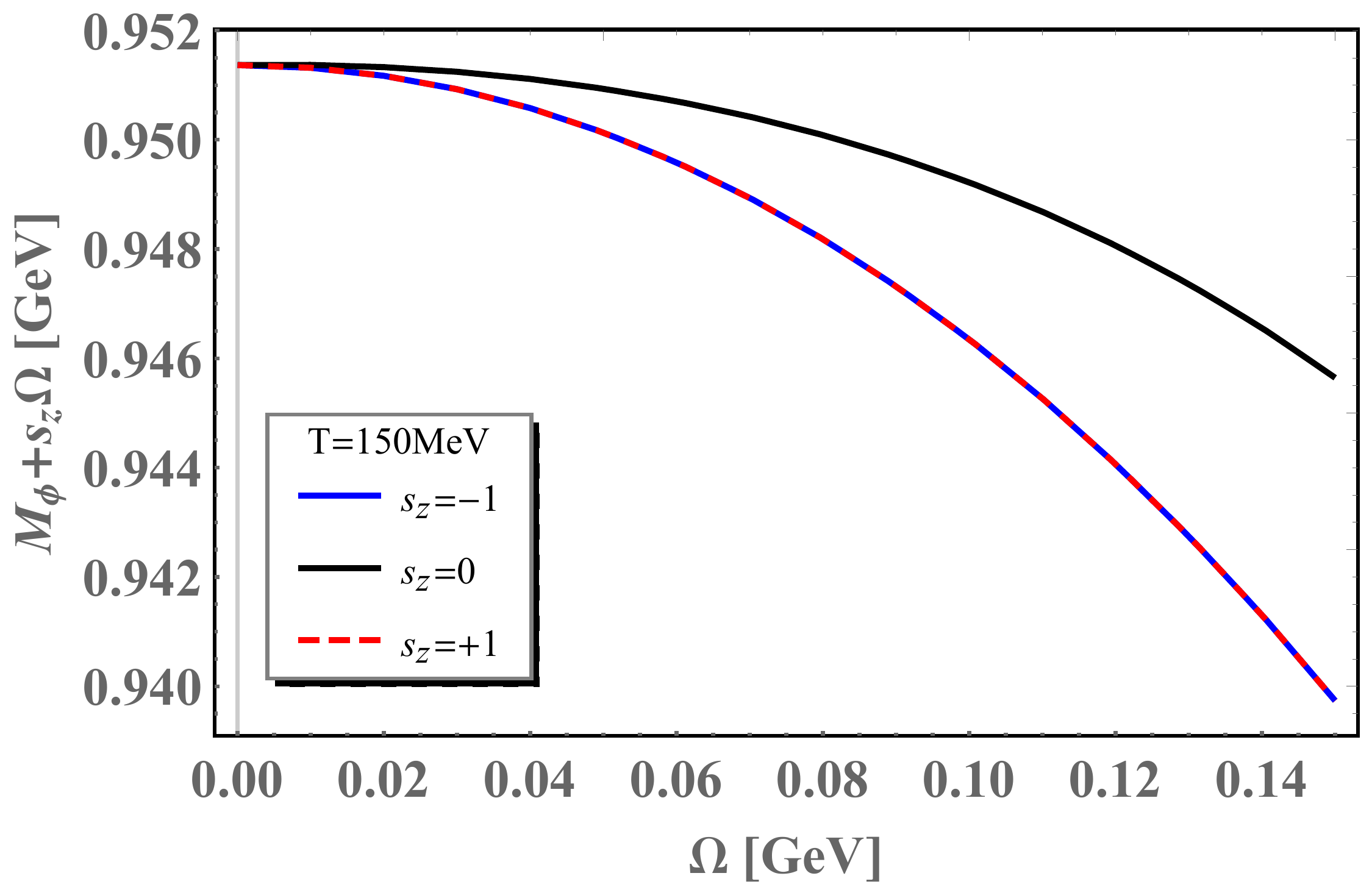}    \caption{The deviation of $\phi$ meson mass as a function of angular velocity at temperature $T=150$ MeV. }  
    \label{fig:masssplit2}
\end{figure}
\begin{figure}[h]
\includegraphics[width=7cm]{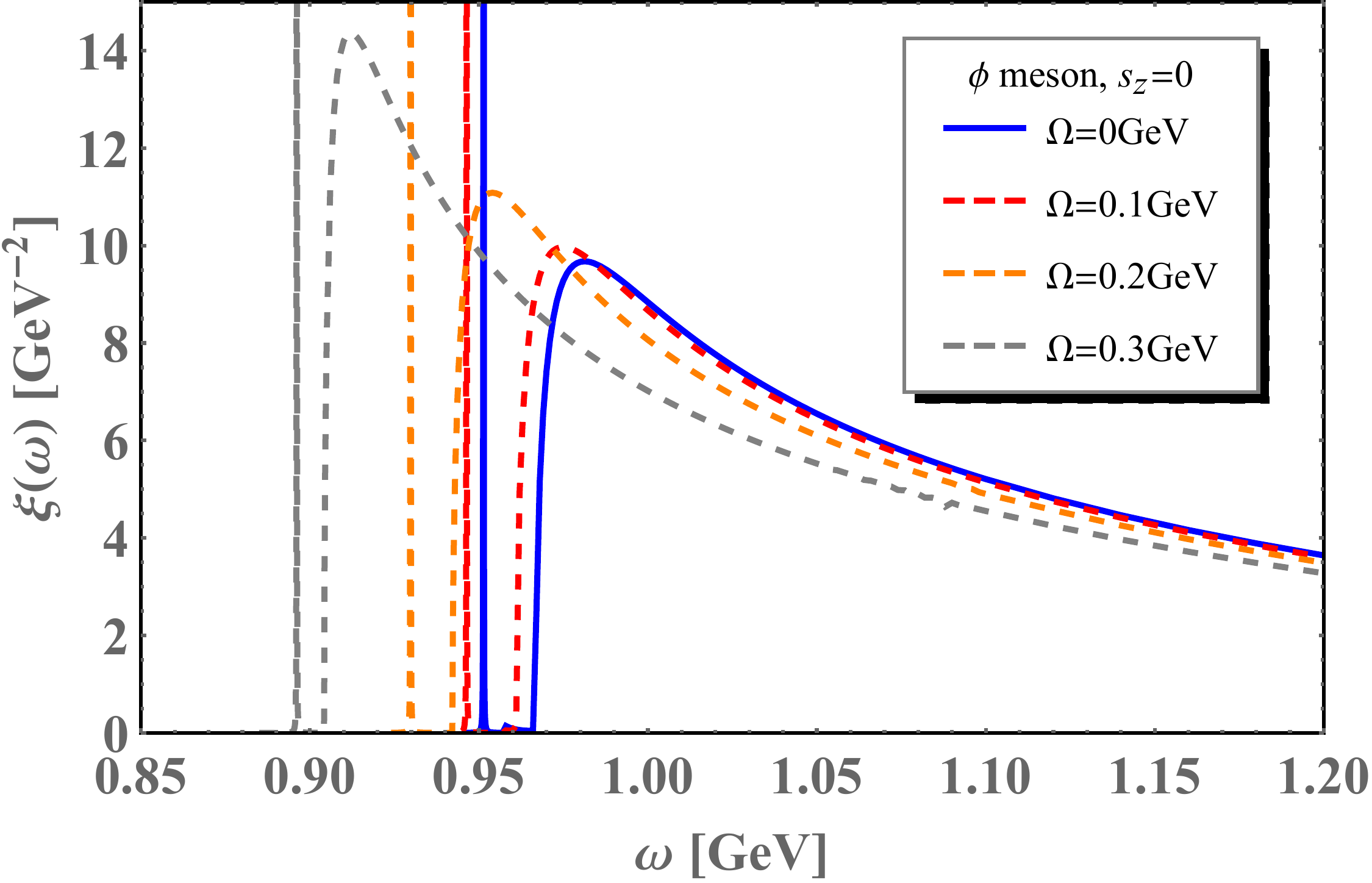}
\includegraphics[width=7cm]{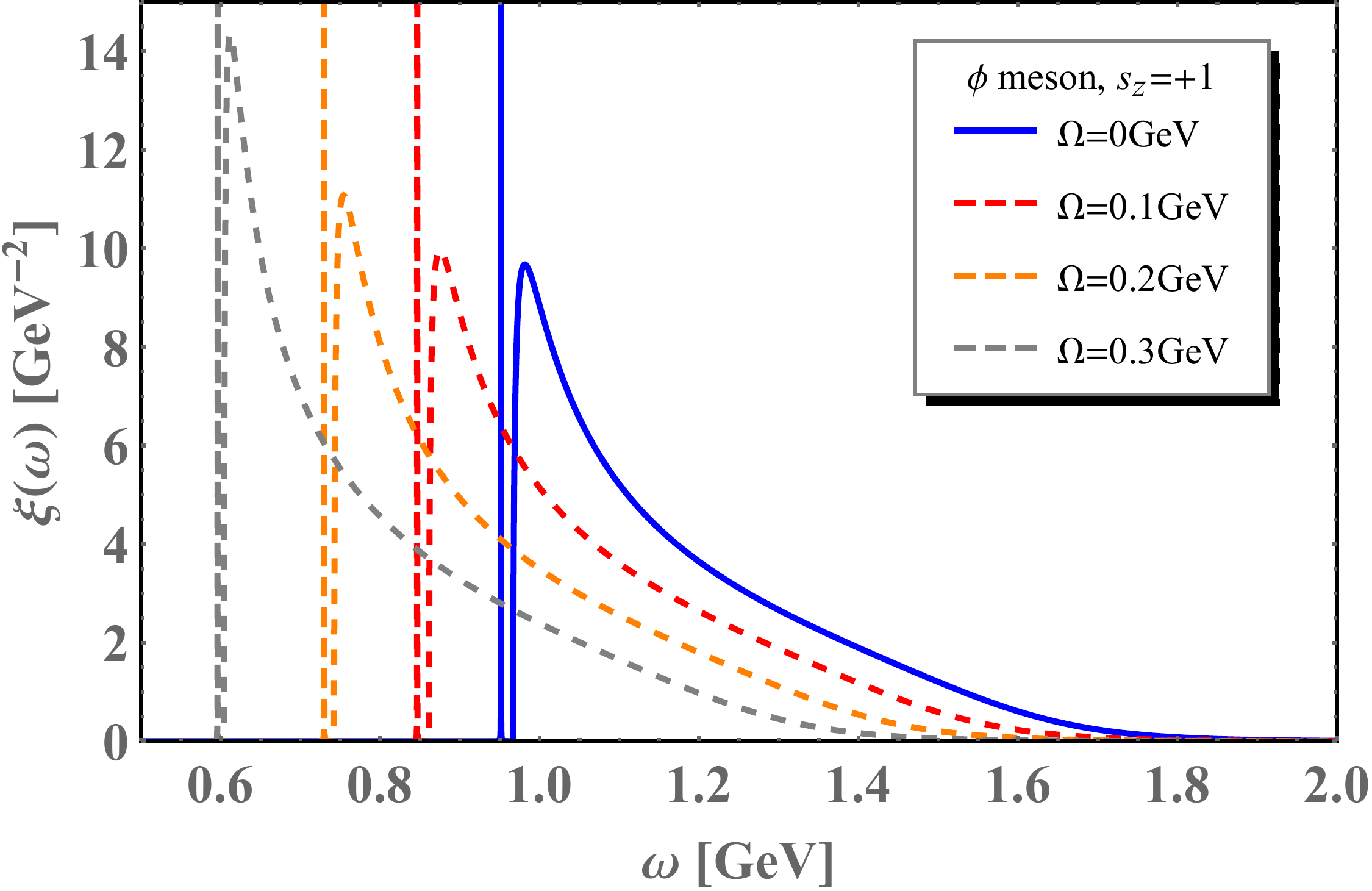}
\includegraphics[width=7cm]{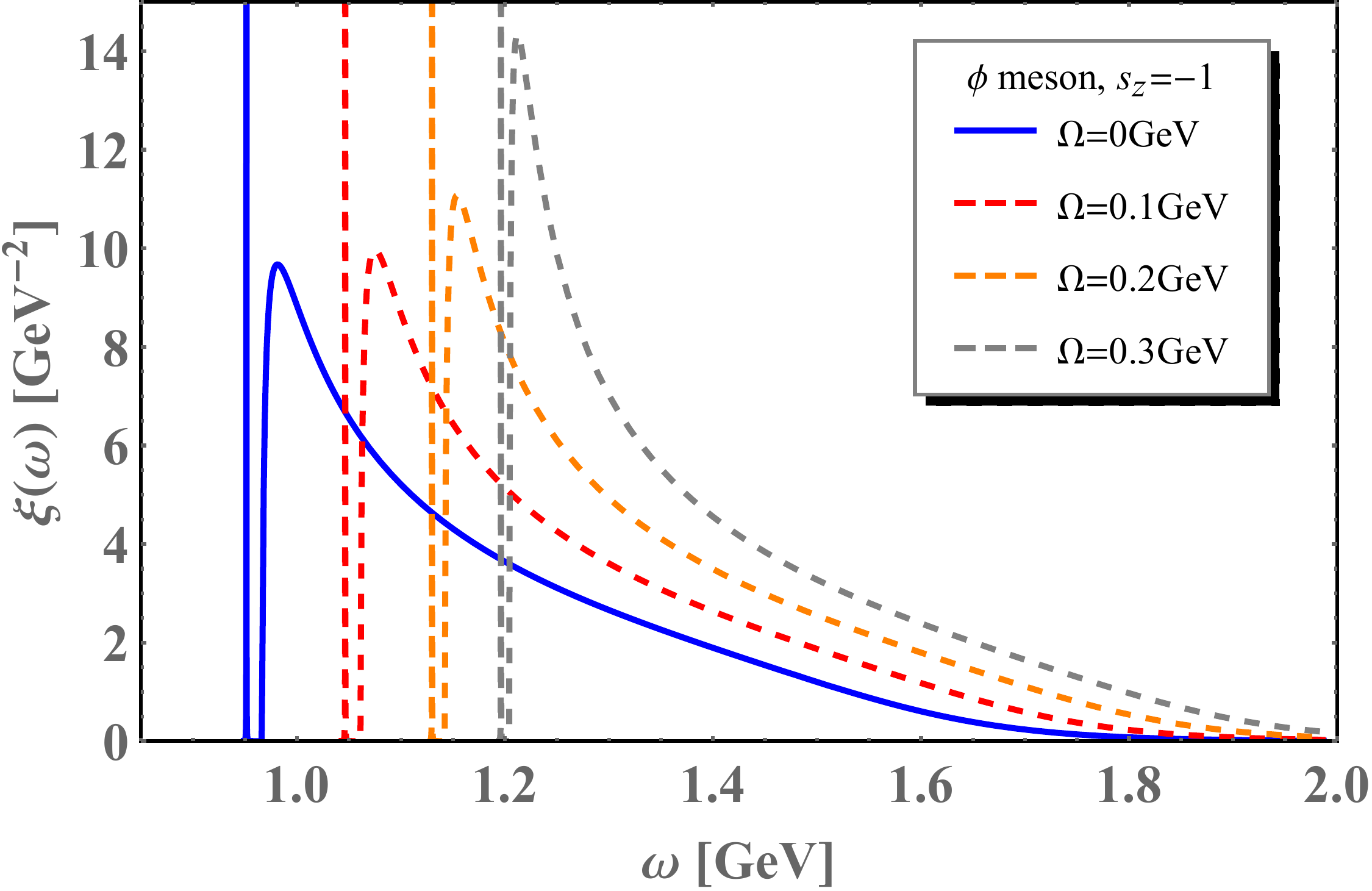}
\caption{\label{fig:massspectra} The spectral function $\xi(\omega)$ for the $\phi$ meson with different spin components as a function of the frequency $\omega$ under different angular velocities $\Omega=0,0.1,0.2,0.3$ GeV at $T=150$ MeV and $\mu=0$.}
\end{figure}
\begin{figure}[h]
\includegraphics[width=7cm]{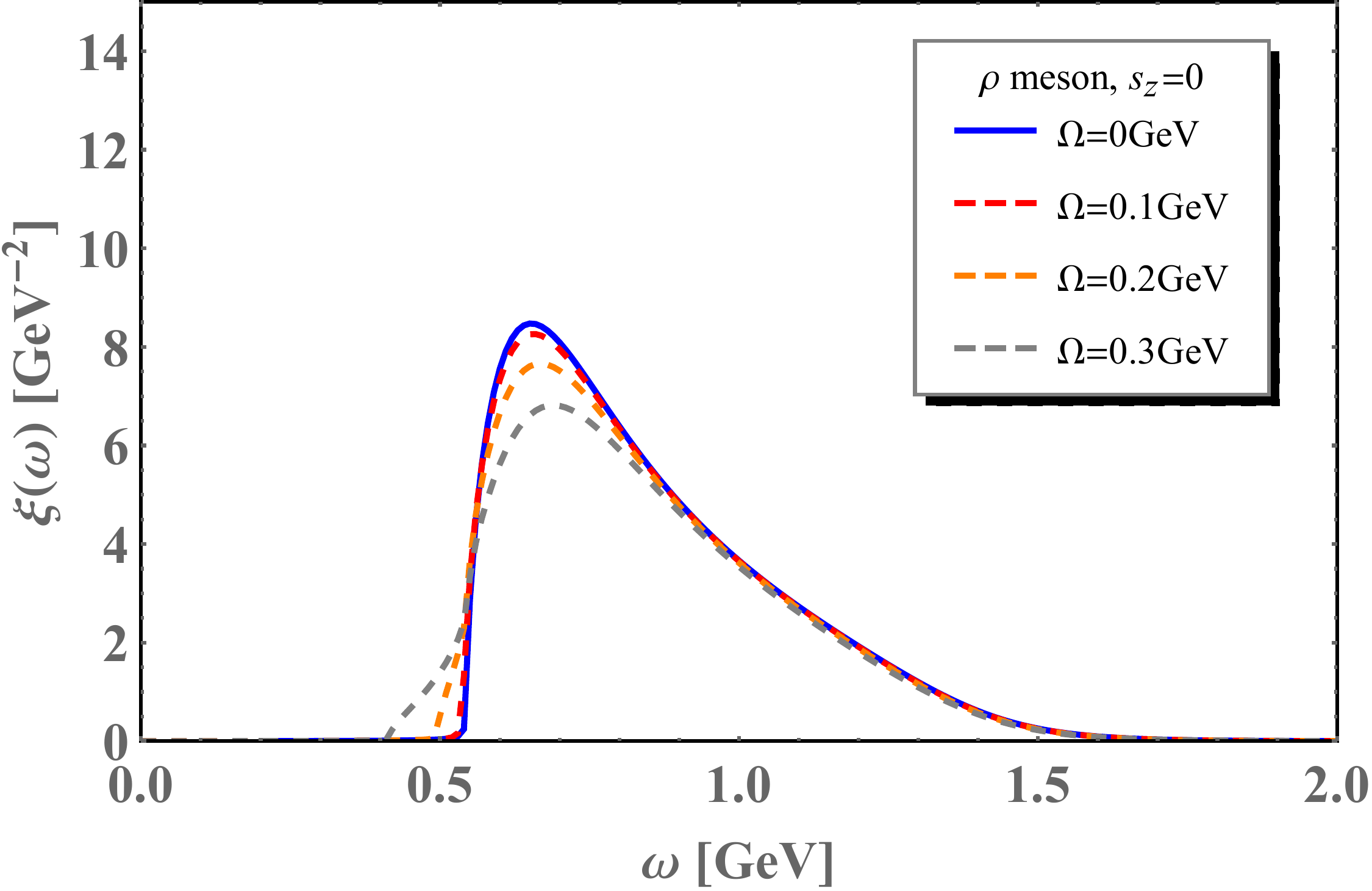}
\includegraphics[width=7cm]{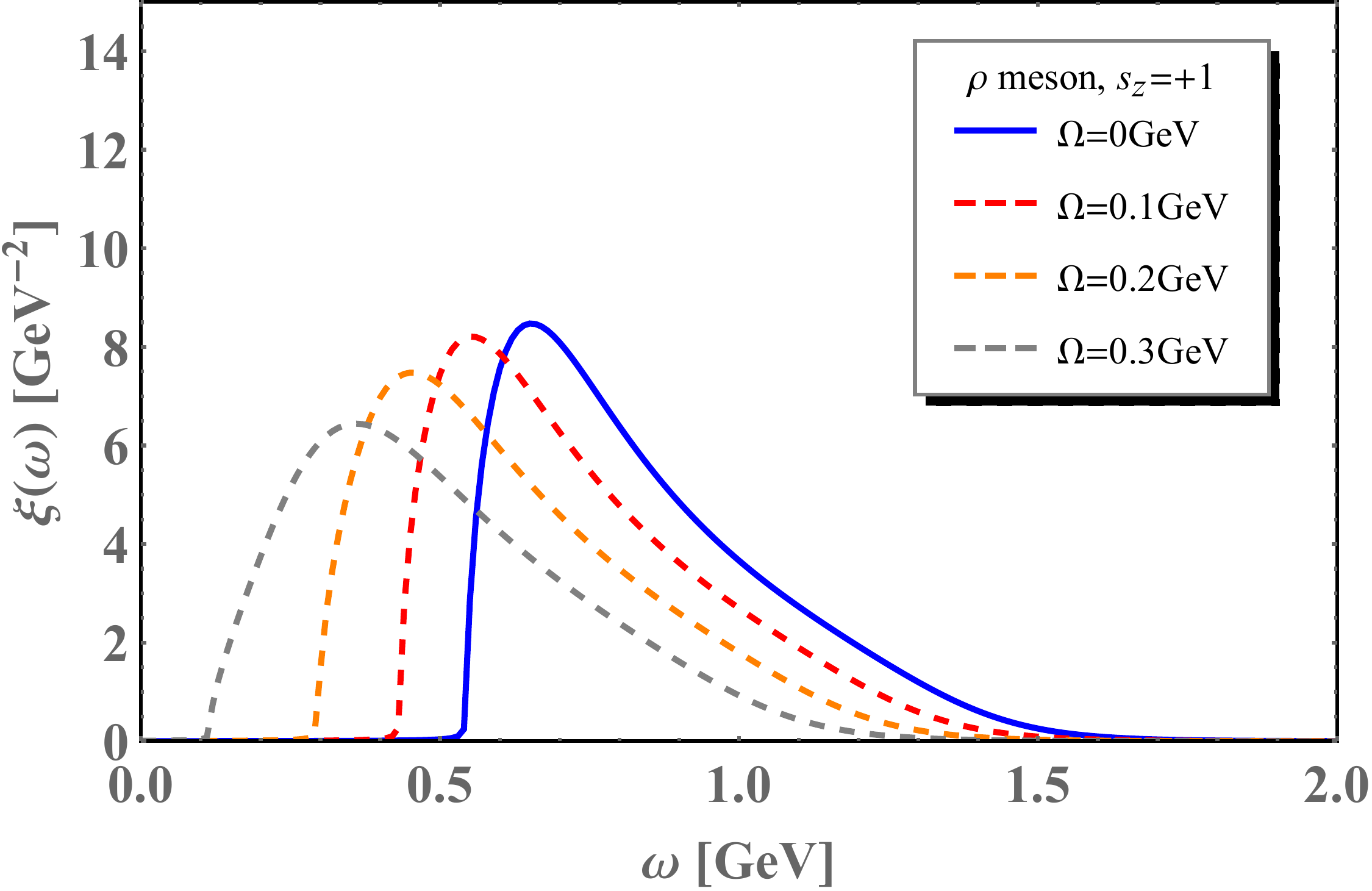}
\includegraphics[width=7cm]{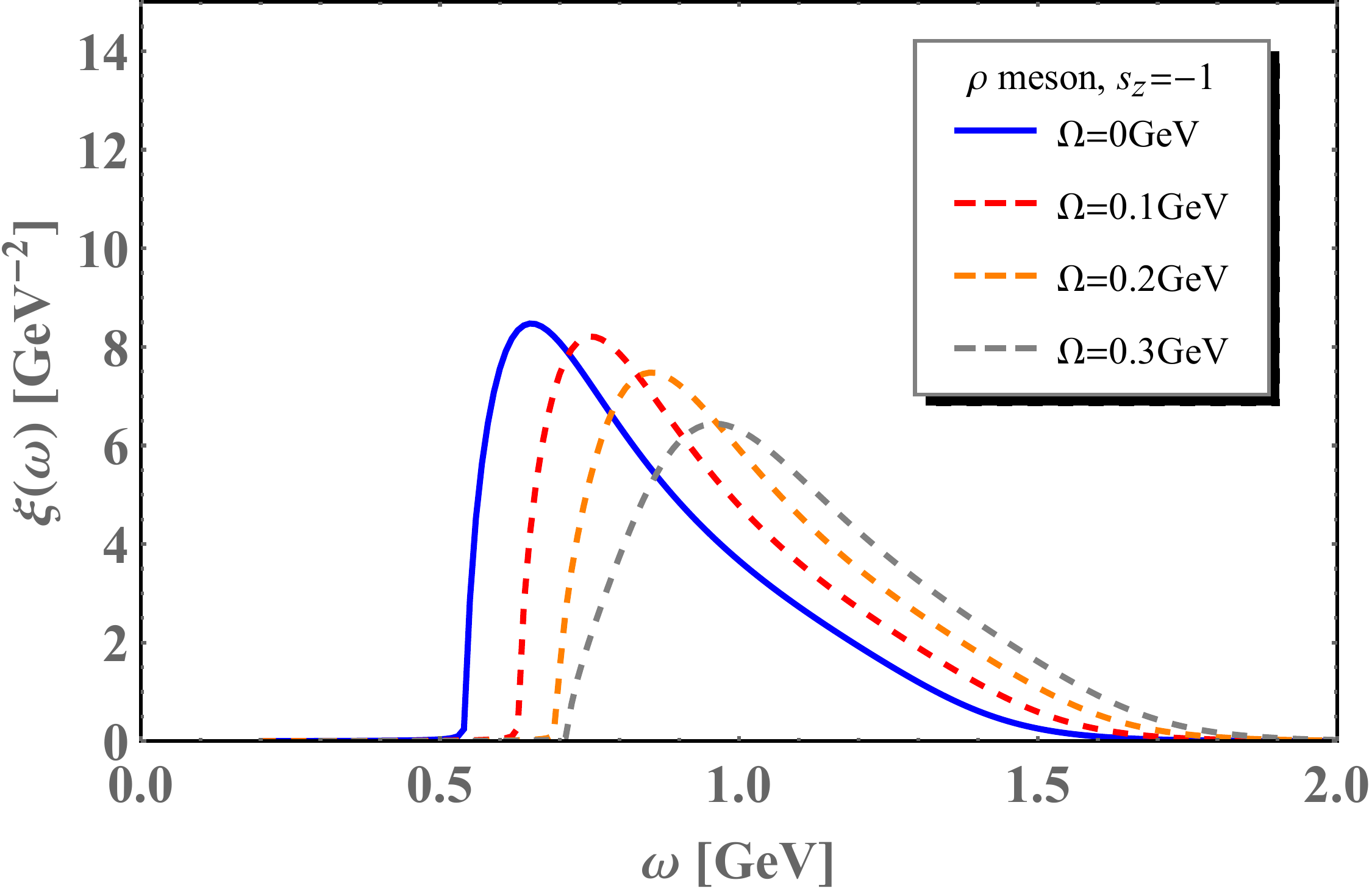}
\caption{\label{fig:rhomassspectra}The spectral function $\xi(\omega)$ for the $\rho$ meson with different spin components as a function of the frequency $\omega$ under different angular velocities $\Omega=0,0.1,0.2,0.3$ GeV  at $T=150$ MeV and $\mu=0$. }
\end{figure}

At $T=150$ MeV and $\mu=0$ MeV, FIG.\ref{fig:masssplit} shows the $\phi$ meson mass with different spin components $s_z=0,\pm1$ as a function of angular velocity. The mass of the $s_{z}=0$ component for the $\phi$ meson remains almost unchanged with the angular velocity. The mass of the $s_{z}=-1$ component of the $\phi$ meson grows almost linearly with the angular velocity. It implies that the $\phi$ meson will be less likely to stay in the $s_{z}=-1$ state. In contrast, the $s_{z}=+1$ component of the $\phi$ meson mass decreases almost linearly with the angular velocity. $s_{z}=+1$ component will be a preferred state under the rotation. Above all, the nearly linear mass splitting behavior of the $\phi$ meson at $T=150$ MeV can be summarized in the following expression:
\begin{equation}
\label{eq:phimassspliting}
  M_\phi(\Omega)\sim M_\phi(0)-s_{z}\Omega.
\end{equation}
In FIG.\ref{fig:masssplit2}, we reveal the deviation of $\phi$ meson mass from Eq.(\ref{eq:phimassspliting}) at $T=150$ MeV. From Eq.(\ref{eq:phimassspliting}), we will know that: $M_{\phi}(\Omega)+s_{z}\Omega\sim M_{\phi}(0)$, and the FIG.\ref{fig:masssplit2} has shown the difference between "$\sim$" and "=". The deviation is caused by the inhibition of chiral condensate. If we expand $M_\phi(\Omega)$ to order $\Omega^2$, the deviation can be revealed as follows:
\begin{equation}
\label{eq:phimassspliting2}
\begin{aligned}
  M_\phi(\Omega,s_{z}=+1)&=0.95-1.00 \Omega-0.54 \Omega^2,\\
  M_\phi(\Omega,s_{z}=0)&=0.95+0.01 \Omega-0.31 \Omega^2,\\
  M_\phi(\Omega,s_{z}=-1)&=0.95+1.00\Omega-0.54 \Omega^2.
\end{aligned}
\end{equation}
Eq.\eqref{eq:phimassspliting2}  shows the deviation of $s_{z}=0,\pm 1$ components, which is caused by the quark mass descending at finite temperature and angular velocity. $\phi$ meson is an almost pure $s\bar{s}$ state~\cite{Klevansky:1992qe}, so its mass is obviously influenced by the quark mass. It is seen that the deviation of $s_{z}=\pm 1$ component states is larger than that of the $s_{z}=0$ component state. 

FIG.\ref{fig:massspectra} and FIG .\ref{fig:rhomassspectra} reveal the rotational effect on the spectral function $\xi(\omega)$ for the $\phi$ meson and $\rho$ meson with different spin components as a function of the frequency $\omega$, respectively. The blue lines stand for spectral functions at $\Omega=0$ without rotation, while the red, orange and gray dashed lines stand for spectral functions under finite rotation in the case of $\Omega=0.1, 0.2$ and $0.3$ GeV, respectively. In the zero rotation case, $\phi$ mesons with different spin states share the same spectral function, which is constituted by a delta function part and a continuum part. The location of the delta function indicates the pole mass. 

For $s_{z}=0$ component of $\phi$ meson, the rotational effect is less remarkable than in the other two cases. So, the scale has been amplified, and we only present the $\xi(\omega)$ in the range of energy $\omega$ from 0.85 GeV to 1.20 GeV. It is found that the spectral functions are shifted slightly to the left. The peaks of the continuum parts are enhanced significantly. 

For the $s_{z}=+1$ component of the $\phi$ meson, the rotational effect will shift the spectral function to the left side, and the rotation will change the height or the shape of the continuum part of the spectral function as well. For the $s_{z}=-1$ component, the spectral function is shifted to the right side correspondingly. 

A similar analysis can be applied for the vector meson $\rho$, since we have assumed $M_{u}=M_{d}$. We can obtain the spectral function by substituting $M_{s}$ for $M_{u}/M_{d}$. For $\phi$ mesons, a bound state is labeled by mass $M_{\phi}$ which is very close to $2 M_{s}$. However, $\rho$ mesons are dissociated at the temperature $T=150$ MeV. So, in FIG.(\ref{fig:rhomassspectra}), a spectral function only has a continuum part and appears as a single peak. The top panel in FIG.(\ref{fig:rhomassspectra}) shows the spectral functions of $\rho$ mesons with spin components $s_{z}=0$. Different colored lines stand for different strengths of angular velocities ranging from $\Omega=0, 0.1, 0.2$ and $0.3$ GeV. Rotational effects are reflected in two aspects: the heights of the peaks are suppressed and the widths are broadened by the angular velocities. It can be understood that mesons tend to be less bounded in a rotating medium. The location of the peaks is almost unchanged in the case of $s_{z}=0$. However, in the case of $s_{z}=+1$, the locations of resonance peaks are shifted to the left side by rotation. Similarly, in the case of $s_{z}=-1$, mass spectra are shifted to the right side by rotation. Above all, on the shape of spectral functions, the rotation effects are similar. 

\subsection{Spin alignment of vector meson $\phi$ and $\rho$}
\label{sec:spinalignmentresult}
\begin{figure}[t]
\centering
\includegraphics[width=7cm]{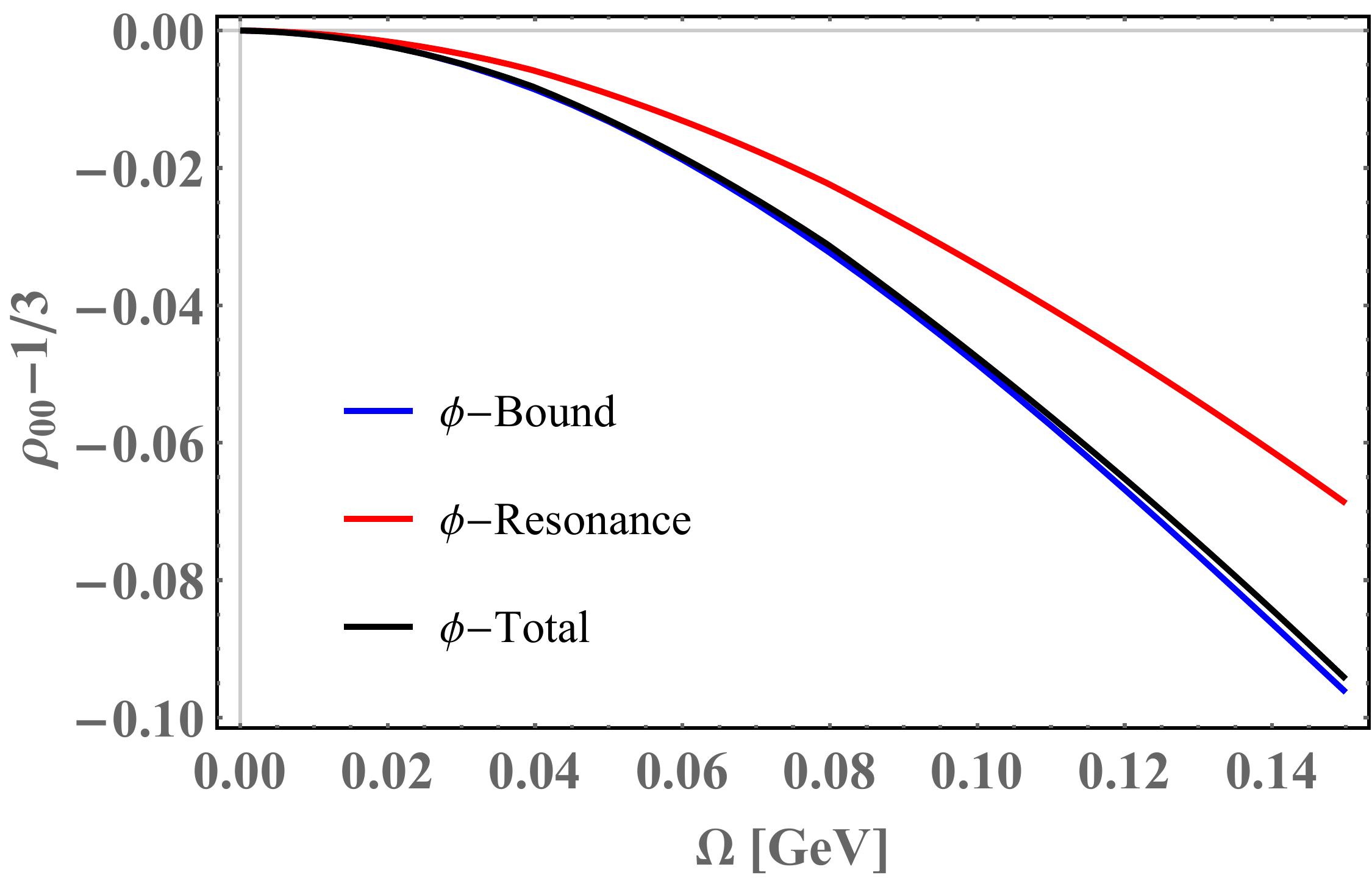} 
\caption{Spin alignment $\rho_{00}$ for vector meson $\phi$ as a function of angular velocity at temperature $T=150$ MeV.}
\label{fig:spinalignment1}
\end{figure}

\begin{figure}[t]
\includegraphics[width=7cm]{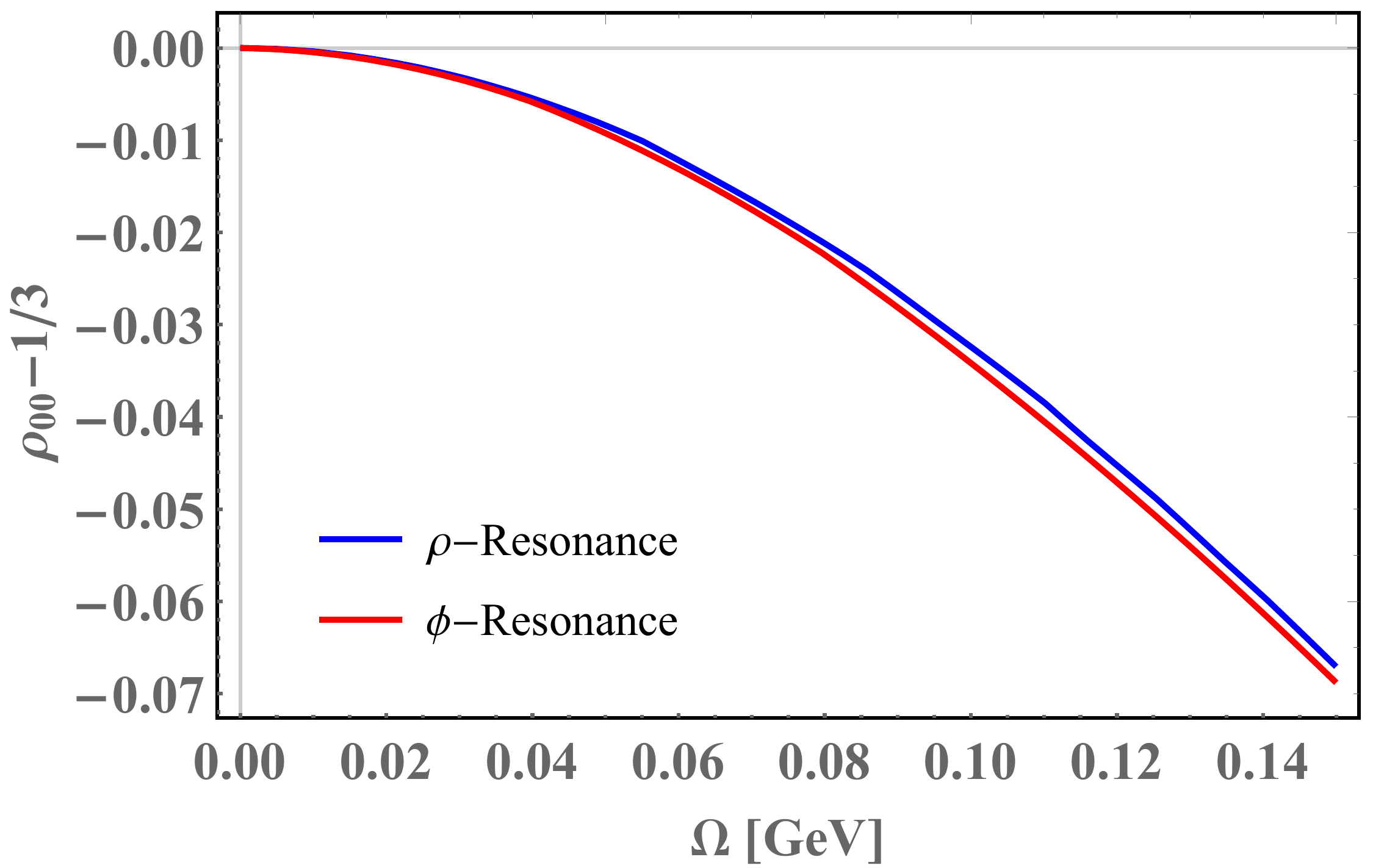} 
\caption{Spin alignment $\rho_{00}$ for resonance states of vector meson $\rho$ and $\phi$ as a function of angular velocity at temperature $T=150$ MeV.}
\label{fig:spinalignment2}
\end{figure}

In FIG.\ref{fig:spinalignment1}, we show the deviation of spin alignment $\rho_{00}$ from 1/3 for the $\phi$ meson as a function of angular velocity at a finite temperature T=150 MeV. In the case of rotation, $\rho_{00}$ is always smaller than 1/3, and the deviation will become more significant as the angular velocity grows. Furthermore, resonance states will have less contribution to spin alignment and the deviation between the bounded state and the total result is even negligible.  In FIG. \ref{fig:spinalignment2}, we compare the spin alignment $\rho_{00}$ from 1/3 for $\phi$ meson with $\rho$ meson, and it is found that the difference is quite small.

Here, we mention the result from the quark coalescence model~\cite{Yang:2017sdk}:
\begin{equation}
\label{eq:fityang}
  \rho_{00}^{\phi}(\Omega)=\frac{1}{3}-\frac{1}{9}(\beta \Omega)^2, 
\end{equation}
where $\beta=1/T$ is the inverse of temperature. Eq.\eqref{eq:fityang} is only valid in the vicinity of $\Omega=0$ GeV, and value of the coefficient is $-\frac{1}{9}\beta^{2}=-4.94 \text{GeV}^{-2}$ for $T=0.15 \text{GeV}$. As a comparison, we fit our result with polynomial functions in a range of angular velocity from $\Omega=0$ GeV to $\Omega=0.15$ GeV. The numerical results give:
\begin{equation}
\rho^{\phi}_{00}(\Omega)=\frac{1}{3}-5.10 \Omega^2+39.62 \Omega^4.
\end{equation}
Here, the dimensions of the angular velocity and the coefficients are omitted. In the vicinity of $\Omega=0$ GeV, the absolute value of the coefficient of $\Omega^{2}$ is larger than $-\frac{1}{9}\beta^{2}=-4.94 \text{GeV}^{-2}$. 

Compared with the spin alignment under an external magnetic field and rotation, the deviation $\rho_{00}-1/3$ is positive under the magnetic field while it is negative in the presence of rotation. It is natural to be understood from quark dynamics that the spin of a particle tends to align along the direction of angular momentum due to the spin-orbital coupling. For the $s_{z}=+1$ component, the vector $\phi,\rho$ meson masses are suppressed in the rotating medium. As a consequence, vector mesons are more possible to occupy the $s_{z}=+1$ state and less possible to occupy $s_{z}=0$ state. So, $\rho_{00}-1/3$ is negative in the rotating medium. On the contrary, $\rho_{00}-1/3$ of $\phi$ meson is positive under the magnetic field~\cite{Sheng:2022ssp}. Actually, vector meson masses in the magnetic field are charge-dependent. The $\phi$ meson is a neutral particle, its property under the magnetic field can be extended from the result of neutral $\rho^0$ meson mass spectra under the magnetic field \cite{Xu:2020sui}. Under the magnetic field, neutral $\phi$ mesons with $s_{z}=\pm 1$ will have a larger mass than $\phi$ mesons with $s_{z}=0$. So, $\phi$ mesons are more possible to occupy the $s_{z}=0$ state in the presence of the magnetic field, which naturally explains why $\rho_{00}-1/3$ for $\phi$ mesons is positive under the magnetic field.

Similarly, our theoretical method can be applied to other species of vector mesons, such as $\rho$ and $K^{*0}$. The difference is the dynamical mass of constituent quark $u$, $d$ and $s$ in the rotating medium. In Eq.\eqref{eq:polfun}, only one species of quark propagator exists in the one-loop polarization function. Since we have applied the assumption $M_{u}=M_{d}$, the spin alignment of the $\rho$ meson is demonstrated in FIG.\ref{fig:spinalignment2}. In the rotating medium with a temperature of 150 MeV, $\rho$ mesons are resonance states. So, we compare it with the resonance states of $\phi$ mesons in FIG.\ref{fig:spinalignment2}. The tendency of the deviation $\rho_{00}-1/3$ is still close to the quadratic polynomial.

It is worth reminding that those results are calculated for vector mesons that stay at rest in a rotating medium, i.e., at $\vec{q}=0$. In the $\vec{q}\neq 0$ case, the calculation will be more complicated, and it is still a puzzle to switch the physical quantities in the rotating frame into the counterparts in the lab frame. Above all, the contribution from the rotating medium is significant, although the spin alignment of vector mesons is affected by a combination of many factors. 

\section{Conclusion and Discussion}
\label{sec:conclusion}
In this work, we investigate the spin alignment of vector mesons $\phi$ and $\rho$ induced by rotation. By applying a three-flavor NJL model with a vector interaction channel, we obtain the dynamical quark mass under rotation. The curves of $M_{f}(\Omega)$ is similar to $M_{f}(\mu)$. For the $s$ quark, the first-order phase transition occurs at a critical angular velocity $\Omega_{c}$, and $M_{s}$ decreases smoothly after the phase transition, which is similar to the quark mass behavior at a finite chemical potential.

After substituting the dynamical quark mass, the mass spectra of vector mesons can be obtained through the quark-antiquark polarization function. The rotating angular velocity induces mass splitting of spin components for vector $\phi,\rho$ mesons $M_{\phi,\rho}(\Omega)\simeq M_{\phi,\rho}(\Omega=0)-s_{z}\Omega$.  This behavior contributes to the spin alignment of vector mesons $\phi,\rho$ in an equilibrium medium. In a rotating medium, $\rho_{00}$ of vector mesons has a negative deviation from $1/3$, which implies a spin alignment phenomenon that can be easily understood from quark dynamics. The spin of a particle tends to align along the direction of angular momentum due to the spin-orbital coupling. For the $s_{z}=+1$ component, the vector meson masses are suppressed in the rotating medium. As a consequence, vector mesons are more possible to occupy the $s_{z}=+1$ state and less possible to occupy the $s_{z}=0$ state. So, $\rho_{00}-1/3$ is negative in the rotating medium.

On the contrary, the deviation $\rho_{00}-1/3$ is positive under the magnetic field, which can also be easily understood from quark dynamics. Under the magnetic field, $s_{z}=\pm 1$ components of $\phi,\rho$ mesons will have a larger mass than that of the $s_{z}=0$ component of $\phi,\rho$ mesons. So, $\phi,\rho$ mesons are more likely to occupy the $s_{z}=0$ state in the presence of the magnetic field, which naturally explains the positive $\rho_{00}-1/3$ for $\phi$ meson under a magnetic field.

Based on a dynamical quark model, in the next step, we are able to calculate the spin alignment $\rho_{00}$ for the transverse momentum-dependent case. Furthermore, we can apply it to the vector meson $K^{*0}$ in the future. In this series of studies, we have studied the vector mesons $\phi$ and $\rho$. In this case, quark loops only contain $s$ and $\bar{s}$ quarks, and light quarks are assumed to have the same value of mass, i.e., $M_{u}=M_{d}$. For the vector meson $K^{*0}$, the mass difference of quarks is expected to explain the different measurements between the vector meson $\phi$ and $K^{*0}$ in the experiment. Our work is an attempt at this target in a rotating medium. 

\begin{acknowledgments}
We thank the helpful discussion with Li Yan, Anping Huang, Xinli Sheng, and Kun Xu. This work is supported in part by the National Natural Science Foundation of China (NSFC) Grant Nos. 12235016, 12221005, the Strategic Priority Research Program of Chinese Academy of Sciences under Grant Nos XDB34030000, the start-up funding from University of Chinese Academy of Sciences(UCAS), and the Fundamental Research Funds for the Central Universities. 
\end{acknowledgments}

\appendix

\section{Quark propagator in a rotating medium}
\label{sec:appendixa}
To obtain the quark propagator in a rotating and dense medium, we adopt the method from Vladimir A. Miransky and Igor A. Shovkovy\cite{Miransky:2015ava}. This derivation has considered the chemical potential $\mu$ and the rotation term $\Omega \cdot \hat{J_z}$. According to an alternative definition, the quark propagator is given by:
\begin{equation}
\small
S\left(\tilde{r}, \tilde{r}^{\prime}\right)=i\left\langle\tilde{r}\left|\left[\left(i \partial_t+\mu+\Omega \cdot \hat{J_z}\right) \gamma^0-\vec{\pi} \cdot \vec{\gamma}-M_{f}\right]^{-1}\right| \tilde{r}^{\prime}\right\rangle \,,
\end{equation}
where $\vec{\pi}$ is the canonical momentum and $\vec{\gamma}$ is the Dirac matrix. Their expressions depend on the coordinates of the position $ \tilde{r}$. At this moment, we treat them as abstract operators, and the propagator can be rewritten as:  
\begin{widetext}
\begin{equation}
\begin{aligned}
S\left(\tilde{r}, \tilde{r}^{\prime}\right)
=&i\left\langle\tilde{r}\left|\left[\left(i \partial_{t}+\mu+\Omega \cdot \hat{J}_z\right) \gamma^{0}-\vec{\pi} \cdot \vec{\gamma}+M_{f}\right] \left[\left(i \partial_{t}+\mu+\Omega \cdot\hat{J}_z\right) \gamma^{0}-\vec{\pi} \cdot \vec{\gamma}+M_{f}\right]^{-1}\right. \right.\\
& {\left.\left.\left[\left(i \partial_{t}+\mu+\Omega \cdot \hat{J_z}\right) \gamma^{0}-\vec{\pi} \cdot \vec{\gamma}-M_{f}\right]^{-1}\right|\tilde{r}^{\prime}\right\rangle} \\
=&i\left\langle\tilde{r}\left|\left[\left(\partial_{t}+\mu+\omega \cdot \hat{J_z}\right) \gamma^{0}-\vec{\pi} \cdot \vec{\gamma}+M_{f}\right] \left[\left(i \partial_{t}+\mu+\Omega\cdot \hat{J}_z\right)^2-\vec{\pi}^2-M_{f}^2\right]^{-1}\right| \tilde{r}^{\prime}\right\rangle \\
&
\end{aligned}
\end{equation}
\end{widetext}
Due to the translation invariance in the $t-$ and $z-$ directions, we can perform the Fourier transformation on the quark propagator as follows:
\begin{equation}
S\left(E, k_{z} ; \mathbf{r}_{\perp}, \mathbf{r}_{\perp}^{\prime}\right)=\int d t d z e^{i E\left(t-t^{\prime}\right)-i k_{z}\left(z-z^{\prime}\right)} S\left(\tilde{r}, \tilde{r}^{\prime}\right).
\end{equation}
Here, $\mathbf{r}_{\perp}=(r,\theta)$ is the position in cylindrical coordinates and $E$ is the energy. The propagator can be expressed as follows:
\begin{equation}
\label{eq:app4}
\begin{aligned}
&S\left(E, k_{z} ; \mathbf{r}_{\perp}, \mathbf{r}_{\perp}^{\prime}\right)\\
= & i\left[\left(E+\mu+\Omega \cdot \hat{J}_{z\left(\mathbf{r}_{\perp}\right)}\right) \gamma^0-\vec{\pi}_{\mathbf{r}_\perp} \cdot \vec{\gamma}_{\perp}-k_{z} \gamma^3+M_{f}\right] \\
& \cdot\left\langle \mathbf{r}_{\perp} \left| \left[\left(E+\mu+\Omega \cdot \hat{J}_z\right)^2-\left(k_{z}\right)^2-\vec{\pi}_{\perp}^2-M_{f}^2\right]^{-1}\right| \mathbf{r}_\perp^{\prime}\right\rangle
\end{aligned}
\end{equation}
where $\vec{\pi}_{\mathbf{r}_\perp}$ and $\hat{J}_{z\left(r_{\perp}\right)}$ are the canonical momentum and the angular momentum operator in cylindrical coordinate space, respectively.
According to Ref.\cite{Jiang:2016wvv}, the operators $\vec{\pi}_\perp^2$ and $\hat{L}_z$ commute with each other and have a common eigenstate $\left|n k_{t}\right\rangle$, which the explicit form in coordinate space is given by Eq.\eqref{eq:statesfermion} and \eqref{eq:statesantifermion}. Here, we present several useful equations:
\begin{equation}
\begin{aligned}
\left\langle\mathbf{r}_{\perp} \mid n k_{t}\right\rangle&=e^{i n \theta} J_n\left(k_t r\right)\\
\hat{L}_z\left.\mid n k_{t}\right\rangle&=n\left.\mid n k_{t}\right\rangle\\
\vec{\pi}_{\perp}\left.\mid n k_{t}\right\rangle&=k_{t}\left.\mid n k_{t}\right\rangle,
\end{aligned}
\end{equation}
where $\hat{L}_z$ is the operator of orbital angular momentum. As a result, the right hand side of Eq.\eqref{eq:app4} can be evaluated as follows:
\begin{widetext}
\begin{equation}
\label{eq:app6}
\begin{aligned}
& \left\langle \mathbf{r}_\perp \left|\left[\left(E+\mu+\Omega \cdot \hat{J}_z\right)^2-\left(k_{z}\right)^2-\vec{\pi}_{\perp}^2-M_{f}^2\right]^{-1} \right|\mathbf{r}_{\perp}^{\prime}\right\rangle \\
= & \sum_n \int k_t d k_t\left\langle \mathbf{r}_{\perp}\left|\left[\left(E+\mu+\Omega \cdot \hat{J}_z\right)^2-\left(k_{z}\right)^2-\vec{\pi}_{\perp}^2-M_{f}^2\right]^{-1}\right| n k_t\right\rangle\left\langle n k_t \mid \mathbf{r}_{\perp}^{\prime}\right\rangle \\
= & \sum_n \int k_{t} d k_t\left\langle \mathbf{r}_{\perp} \left|\left[\left(E+\mu+\Omega \cdot \hat{L}_z\right)^2+2(E+\mu) \Omega \cdot  S_{z}+\Omega^2\left(S_{z}\right)^2-k_z^2-\vec{\pi}_{\perp}^2-M_{f}^2\right]^{-1} \right|n k_{t}\right\rangle\left\langle n k_t \mid \mathbf{r}_{\perp}^{\prime}\right\rangle\\
= & \sum_n \int_0^{+\infty} k_t d k_t J_n\left(k_t r\right) J_n\left(k_t r^{\prime}\right) e^{i n\left( \theta-\theta^{\prime}\right)}\left[(E+\mu+\Omega n)^2+2(E+\mu)\Omega S_{z}+\Omega^2 \cdot \frac{1}{4}-k_z^2-k_t^2-M_{f}^2\right]^{-1}
\end{aligned}
\end{equation}
\end{widetext}
where $S_{z}=\frac{i}{2} \gamma^{1}\gamma^{2}$ is the spin angular momentum term. And we have inserted the completeness condition in the second line of Eq.\eqref{eq:app6}. Now, it is easy to obtain Eq.\eqref{eq:propagator} by means of projection operator $\mathcal{P}_{\pm}=\frac{1}{2}\left(1 \pm i \gamma^1 \gamma^2\right)$. By inserting $I_4=\mathcal{P}_{-}+\mathcal{P}_{+}$, the summation in Eq.\eqref{eq:app6} can be replaced by:
\begin{widetext}
\begin{equation}
\label{eq:app7}
\begin{aligned}
\sum_n  \frac{J_n\left(k_t r\right) J_n\left(k_t r^{\prime}\right) e^{i n\left( \theta-\theta^{\prime}\right)}\mathcal{P}_{+}+
J_{n+1}\left(k_t r\right) J_{n+1}\left(k_t r^{\prime}\right) e^{i (n+1)\left( \theta-\theta^{\prime}\right)}\mathcal{P}_{-}}{\left[E+\mu+\left(n+\frac{1}{2}\right) \Omega\right]^2-k_z^2-k_t^2-M_{f}^2}.
\end{aligned}
\end{equation}
\end{widetext}
Finally, we can calculate Eq.\eqref{eq:app4} by acting the operators $ \hat{J}_{z\left(\mathbf{r}_{\perp}\right)} \gamma^0$, $\vec{\pi}_{\mathbf{r}_\perp} \cdot \vec{r}_{\perp}$ and $k_{z} \gamma^3$ on Eq.\eqref{eq:app7}.
\section{Spectral functions}
\label{sec:appB}
The explicit form and the derivation of spectral functions can be found in Ref.\cite{WeiMingHua:2020eee,Wei:2021dib}. In our previous work, we have utilized the properties of Bessel functions, i.e., $J_{n}(0)=0$ for $n \neq 0$ and $J_{0}(0)=1$. As a consequence, the infinite summation is reduced to finite terms. Then, Eq.\eqref{eq:polfun} can be evaluated at zero and finite temperature. For saving space in this manuscript, we merely present one component of polarization functions. For example, the imaginary part of the 00-component is~\cite{Wei:2021dib}:
\begin{widetext}
\begin{equation}
\begin{aligned}
\operatorname{Im} \Pi^{00}(\omega, \vec{q})= & -\frac{\pi}{2} N_f N_c \sum_{\eta=\pm 1} \int \frac{d^3 \vec{p}}{(2 \pi)^3} \frac{1}{E_p E_k}\left\{\left[E_p E_k+\vec{p} \cdot \vec{k}+M_f^2\right]\left[f\left(E_p-\mu-\frac{\eta \Omega}{2}\right)+f\left(E_p+\mu-\frac{\eta \Omega}{2}\right)\right]\right. \\
& \times\left[\delta\left(\omega+E_p-E_k\right)-\delta\left(\omega-E_p+E_k\right)\right]+\left[E_p E_k-\vec{p} \cdot \vec{k}-M_f^2\right] \delta\left(\omega-E_p-E_k\right) \\
& \left.\times\left[1-f\left(E_p-\mu-\frac{\eta \Omega}{2}\right)-f\left(E_p+\mu-\frac{\eta \Omega}{2}\right)\right]\right\}.
\end{aligned}
\end{equation}
\end{widetext}
where $k=p+q$ and $E_{k}=\sqrt{\vec{k}^{2}+M_{f}^{2}}$. Here, $N_{f}$ and $N_{c}$ are the flavor number and color number in the quark loop, and $f(x)$ is the Fermi-Dirac distribution function with a finite temperature $T$. Other components can be obtained similarly.
\nocite{*}

\bibliography{apssamp}

\end{document}